\documentclass[letterpaper]{article}

\usepackage{bm}
\usepackage{graphicx,fancyhdr,lscape,dsfont,amsmath,amssymb} 
\usepackage{color, amsthm}
\usepackage{longtable}
\usepackage{arydshln}
\usepackage{psfrag}
\usepackage{subfigure} 
\usepackage{multirow}
\usepackage{stmaryrd}
\usepackage[titletoc]{appendix}
\usepackage{natbib}
\usepackage{ulem} 
\usepackage{mathrsfs}
\usepackage{bbm}
\usepackage{upgreek}
\usepackage{epstopdf}
\usepackage{booktabs}

\usepackage[T1]{fontenc}
\usepackage[latin9]{inputenc}
\usepackage{geometry}
\usepackage{natbib}
\usepackage[usenames,dvipsnames]{xcolor}
\usepackage{esint}
\usepackage{soul}
\definecolor{lightyellow}{rgb}{1.0, 1.0, 0.88}
\definecolor{lightmauve}{rgb}{0.86, 0.82, 1.0}
\sethlcolor{lightmauve}
\usepackage{framed}
\usepackage{fancybox}
\colorlet{shadecolor}{lightmauve}

\newcommand{\trace}[1]{ {\rm tr} \left[ \, {#1} \, \right] }

\usepackage{graphics}
\usepackage[affil-it,auth-sc]{authblk}
\usepackage{array}

\newcolumntype{L}[1]{>{\raggedright\let\newline\\\arraybackslash\hspace{0pt}}m{#1}}
\newcolumntype{C}[1]{>{\centering\let\newline\\\arraybackslash\hspace{0pt}}m{#1}}
\newcolumntype{R}[1]{>{\raggedleft\let\newline\\\arraybackslash\hspace{0pt}}m{#1}}

\newcommand{\bge}{\TextOrMath{$\boldsymbol{\varepsilon}$\xspace}{\boldsymbol\varepsilon}}
\newcommand{\bga}{\TextOrMath{$\boldsymbol{\alpha}$\xspace}{\boldsymbol\alpha}}
\newcommand{\bgb}{\TextOrMath{$\boldsymbol{\beta}$\xspace}{\boldsymbol\beta}}
\newcommand{\bgg}{\TextOrMath{$\boldsymbol{\gamma}$\xspace}{\boldsymbol\gamma}}

\newcommand{\bgG}{\TextOrMath{$\boldsymbol{\Gamma}$\xspace}{\boldsymbol\Gamma}}
\newcommand{\bgO}{\TextOrMath{$\boldsymbol{\Omega}$\xspace}{\boldsymbol\Omega}}

\allowdisplaybreaks

\newcommand{\beq}{\begin{equation}}
\newcommand{\eeq}{\end{equation}}

\newcommand{\beqar}{\begin{eqnarray}}
\newcommand{\eeqar}{\end{eqnarray}}
\newcommand{\bit}{\begin{itemize}}
\newcommand{\eit}{\end{itemize}}
\newcommand{\benum}{\begin{enumerate}}
\newcommand{\eenum}{\end{enumerate}}
\newcommand{\barr}{\begin{array}}
\newcommand{\earr}{\end{array}}

\newcommand\eq[1]{(\ref{#1})}
\newcommand{\bfm}[1]{\mbox{\boldmath ${#1}$}}        

\def\XXint#1#2#3{{\setbox0=\hbox{$#1{#2#3}{\int}$}
   \vcenter{\hbox{$#2#3$}}\kern-.5\wd0}}

\def\b0{\mbox{\boldmath $0$}}

\def\be{\mbox{\boldmath $e$}}

\def\bj{\mbox{\boldmath $j$}}

\def\bq{\mbox{\boldmath $q$}}

\def\bu{\mbox{\boldmath $u$}}

\def\bx{\mbox{\boldmath $x$}}

\def\bA{\mbox{\boldmath $A$}}
\def\bB{\mbox{\boldmath $B$}}
\def\bC{\mbox{\boldmath $C$}}

\def\f0{\ensuremath{\mathbb{O}}}

\newcommand{\BGs}{\bfm\sigma}

\newcommand{\mD}{\ensuremath{\mathcal{D}}}

\title{The generalized Floquet-Bloch spectrum \\ for periodic thermodiffusive layered materials}

\author{F. Fantoni$^{1*}$, L. Morini$^{2}$,  A. Bacigalupo$^{3}$\footnote{Corresponding authors: Tel:+39 0303711330, $\hspace{10cm}$ 
 E-mail addresses: francesca.fantoni@unibs.it; andrea.bacigalupo@unige.it}
 , M. Paggi$^4$
\\
\begin{small}
$^{1}$ DICATAM, Universit\`a degli Studi di Brescia, via Branze 43, 25123, Brescia, Italy
\end{small}
\\
\begin{small}
$^{2}$School of Engineering, Cardiff University, Cardiff, CF24 3AA, United Kingdom
\end{small}
\\
\begin{small}
$^{3}$ DICCA, Universit\`a degli Studi di Genova,via Montallegro 1,
16145 Genova, Italy
 \end{small}
 \\
 \begin{small}
 $^4$IMT School for Advanced Studies Lucca, Piazza San Francesco 19, Lucca, 55100, Italy
 \end{small}
}

\makeatother
\begin{document}


\maketitle 
\begin{abstract}
The dynamic behaviour of periodic thermodiffusive multi-layered media excited by harmonic oscillations is studied. In the framework of linear thermodiffusive elasticity, periodic laminates, whose elementary cell is composed by an arbitrary number of layers, are considered. The generalized Floquet-Bloch conditions are imposed, and the universal dispersion relation of the composite is obtained by means of an approach based on the formal solution for a single layer together with the transfer matrix method. The eigenvalue problem associated with the dispersion equation is solved by means of an analytical  procedure based on the symplecticity properties of the transfer matrix to which corresponds a palindromic characteristic polynomial, and the frequency band structure associated to wave propagating inside the medium are finally derived. The proposed approach is tested through illustrative examples where thermodiffusive multilayered structures of interest for renewable energy devices fabrication are analyzed. The effects of  thermodiffusion coupling on both the propagation and attenuation of Bloch waves in these systems are investigated in detail.\\

\end{abstract}

\section{Introduction}
\label{Sec::Introduction}
In the last years, composite laminates subject to thermodiffusive phenomena have been largely used in the design and fabrication of renewable energy devices characterized by a multi-layered configuration, such as lithium-ion batteries \citep{Ellis1,Salvadori1}, solid oxide fuel cells (SOFCs) \citep{Kakac2007,Colpan2008,Kim2009,Kuebler2010,Hasanov2011,Nakajo1, Dev1} and photovoltaic modules (PV) \citep{Paggi1}. Several studies \citep{Atkinson1, Delette1} have shown that, in real operative scenarios,
performances in terms of power generation and energy conversion efficiency can be compromised because of the severe thermomechanical stress as well as intense particle flows to which components of such energy devices are subjected \citep{Muramatsu2015}. This can ultimately impact on their resistance to damage with resulting cracks formation and spreading. Consequently, modeling and predicting these phenomena is a crucial issue in order to ensure the successful manufacture of multi-layered renewable energy devices and to optimize their performances.
Energy devices of this kind are generally organized in stacks where more elements are separated by metallic interconnections \citep{Molla2016}. Due to their particular structure, they can be modelled as periodic themodiffusive laminates which elementary cell, representing the single device, is composed of an arbitrary number of elasto-themodiffusive phases. {This idealised representation provides the possibility of estimating the overall mechanical and thermodiffusive properties of such 
multi-layered systems through homogenization methods, avoiding the challenging computations required by the direct numerical study of the heterogeneous structures  \citep{Bove1,Richard1, Haji1}.
Homogenization techniques, in fact, allow to take into account the role of the microstructure upon the overall constitutive behaviour  of composite materials in a concise, but accurate way.
They have been a matter of extremely intensive research within the last decades and, in a general sense, homogenization procedures can be classified in asymptotic techniques \citep{BakhPan1} togheter with their extension to multi-field phenomena \citep{fantoni2017, fantoni2018,Fantoni2019phase},
variational-asymptotic techniques \citep{SmiCher1}, and different identification approaches, involving the analytical \citep{BigDru1,Bacca1,Bacca2} and computational techniques 
\citep{Forest3,Lew2004,Scarpa2009,DeBellis1,Forest2,Wang2017,Yvonnet2020}.
Furthermore, dynamic homogenization schemes, useful to approximate frequency band structure of periodic media at high frequencies, can be found in \citep{Zhikov2000,Smyshlyaev2009,Craster2010,Bacigalupo2016high,Sridhar2018,
Kamotski2019,Bacigalupo2019}.
In the context of multi-field asymptotic homogenization methods applied to thermodiffusive phenomena, \cite{Bacigalupo8,Bacigalupo9} investigated the static overall constitutive properties of periodic media.
The dynamics of periodic thermodiffusive devices has been subsequently investigated via asymptotic homogenization in \cite{fantoni2020}.
The dynamic behaviour of laminate media with microstructure has been extensively studied   \citep{Qian2004,Willis2009,Nemat2011,Caviglia2012}. 
In the context of energy devices, the interest
 is motivated by the fact that  media are subject to intrinsically dynamic phenomena such as shock thermal waves, interface waves and instabilities which cannot be described in the framework of static, quasi-static or steady-state formulations. In this regard, an accurate analysis of the harmonic waves propagation in thermodiffusive laminate media, and especially of the effects of the coupling between mechanical, thermal and diffusive observables,} has not been addressed in details to the authors knowledge. Indeed, most of the work conducted on this topics have been performed adopting the generalized theories of thermoelasticity 
and thermodiffusion \citep{Lord1, Sherief1}. 
{These approaches provide thermal and diffusive relaxation times and then the standard heat and mass conduction equations are transformed in hyperbolic type equations. 
In doing so,
 both the temperature and the mass fields evolve in the medium in form of heat and diffusive waves having finite propagation speeds which interact with the mechanical waves. 
 In contrast, we propose a different formulation assuming
 that the elastic waves equation is coupled with the standard heat conduction and mass diffusion equations. 
 These lasts  are of parabolic type and are then associated with an imaginary part of the spectrum corresponding to damping phenomena.}
  We implement the generalized Floquet-Bloch quasiperiodic conditions, and by means of a generalization of the transfer matrix method \citep{Hawwa1}, 
we derive a general expression for the characteristic equation 
valid for periodic thermodiffusive laminates which elementary cell is composed by an arbitrary number of phases. 
{The transfer matrix method has been widely exploited in order to investigate waves propagation  in periodic media \citep{Adams2008,Shmuel2016,Lee2017,Wang2018}.}
Symplecticity properties of the transfer matrix to which corresponds a palindromic characteristic polynomial are exploited in order to solve the eigenvalues problem associated with the characteristic equation, and this general procedure provides the frequency band structure ({\itshape complex  spectra}) associated to wave propagating inside the medium. The potentialities of this technique are illustrated through illustrative examples where the propagation and damping of harmonic thermal and diffusive oscillations as well as of mechanical waves in bi-phase laminates of interest for SOFCs realization is addressed. {The observed damping effects 
are due to the imaginary part of the spectra, which derives from the parabolicity of heat conduction and mass transfer equations. Therefore, these phenomena cannot be detected by means of the other approaches currently available in the literature, based on hyperbolic equations.} 
The paper is organized as follows: Section 2 summarizes governing equations for a linear thermodiffusive material and the wave-like expression of harmonic plane oscillations propagating inside the medium. 
Section 3 is dedicated to present the generalization of the transfer matrix method exploited to obtain, together with Floquet-Bloch conditions,  complex spectra for thermodiffusive laminates.
Representative examples are performed in Section 4, thus showing complex spectra obtained for bi-phase isotropic thermodiffusive laminates of interest for SOFCs fabrication in order to investigate the effects of thermodiffusive coupling upon propagation and damping properties of elastic waves traveling inside the composite.
Finally, conclusions are addressed in Section 5.

\section{Problem formulation}
\label{Sec::ProblemFormulation}
\begin{figure}[!htb]
\centering
\begin{minipage}[c]{.49\textwidth}
\centering
\includegraphics[scale=0.46]{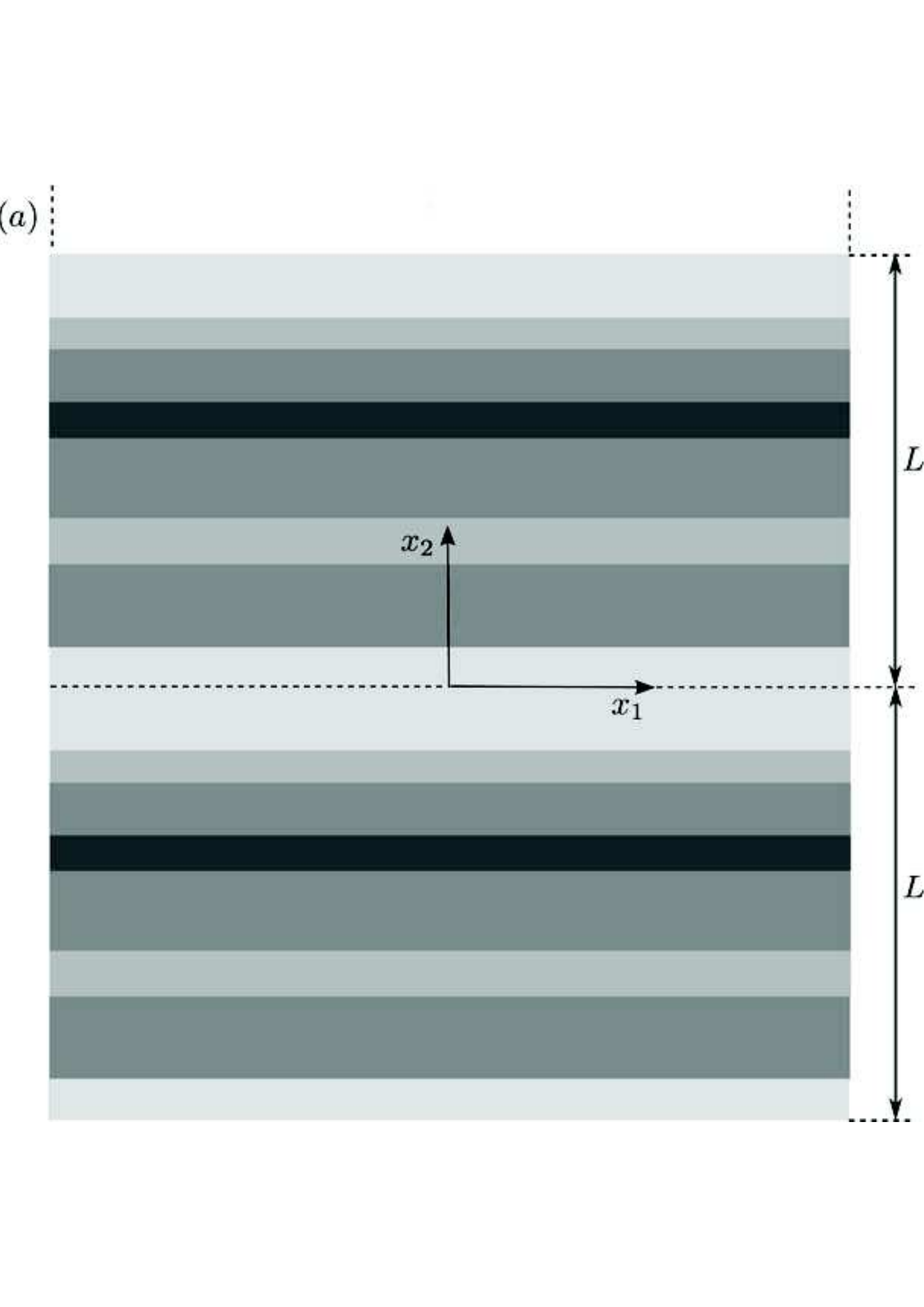}
 \end{minipage}
 \begin{minipage}[c]{.49\textwidth}
 \centering
\includegraphics[scale=0.46, angle=-90]{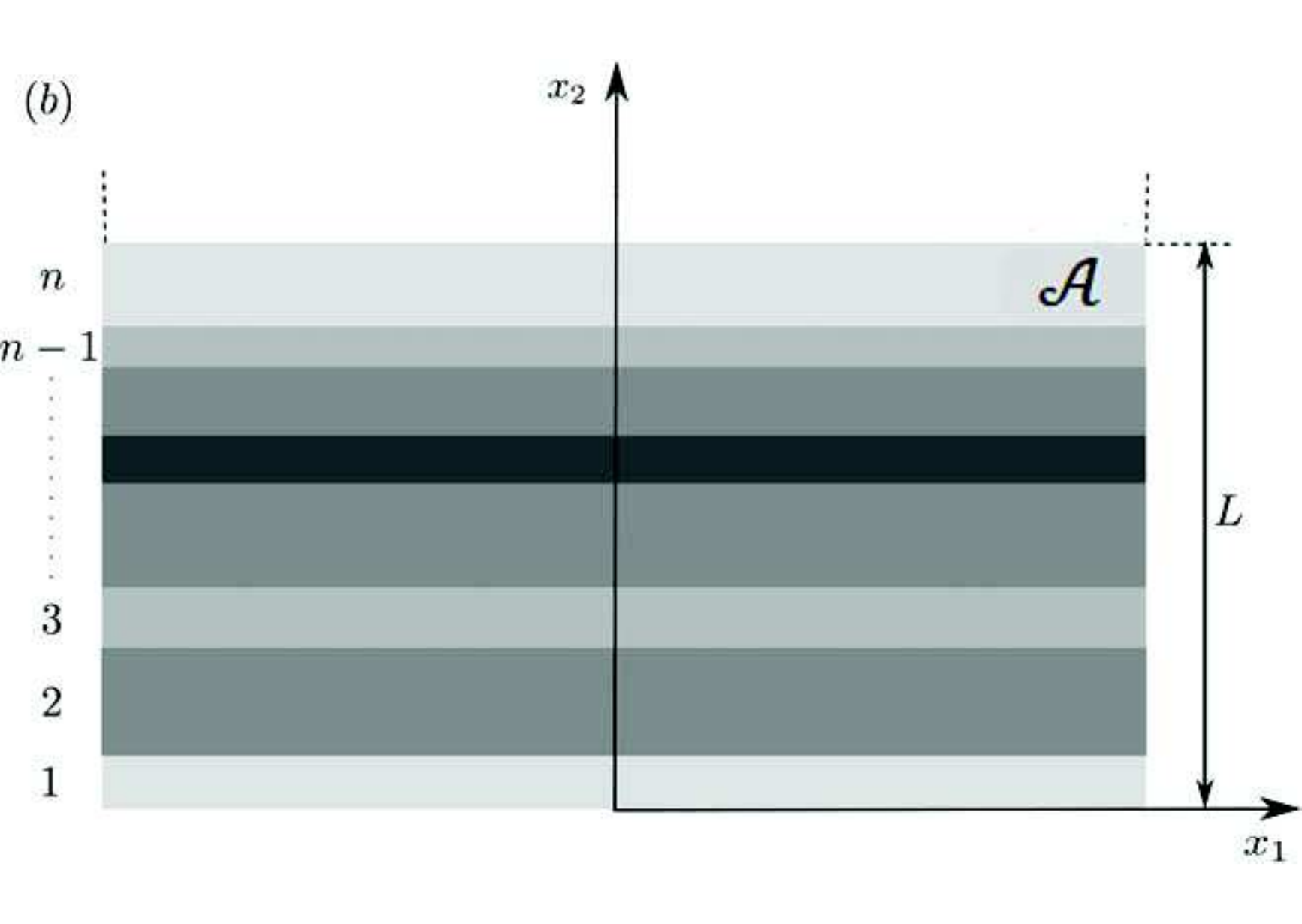}
 \end{minipage}
 \protect\protect\caption{(a) Periodic thermodiffusive laminate; 
(b) Periodic cell $\mathcal{A}$ composed by $n$ layers of arbitrary thickness. \label{thermolam}}
\end{figure}

One considers a plane thermodiffusive laminate medium whose periodic cell is composed by an arbitrary number of layers $n$ perfectly bonded at their interfaces and stacked along the $x_{2}-$axis (see figure \ref{thermolam} ). 
Each material point is identified by the position vector $\bx=x_{1}\be_{1}+x_{2}\be_{2}$ referred to a system of coordinates with origin at point $O$ and orthogonal base $[\be_1, \be_2]$. {The periodic cell
$\mathcal{A}$ has a characteristic length equal to $L$ in the direction perpendicular to material layering and it is translationally invariant along the layering.
As depicted in figure \ref{thermolam}-(b),  $L=\sum_{m=1}^{n}\ell_{m}$ where $\ell_{m}$ represents the thickness of each single layer. 
In the followings, governing 
equations  for a linear thermodiffusive material are introduced.}
%
\subsection{{Governing equations for a linear thermodiffusive material}}
\label{governing}
Assuming that the constituent layers of the laminate are  linear thermodiffusive elastic media,
the three fields characterizing the behaviour of the thermodiffusive material are the displacement $\mathbf{u}(\mathbf{x},t)=u_i(\mathbf{x},t)\mathbf{e}_i$, the relative temperature $\theta(\mathbf{x},t)=T(\mathbf{x},t)-T_0$, with $T_0$ the temperature of the natural state, and the relative chemical potential $\eta(\mathbf{x},t)=P(\mathbf{x},t)-P_0$ with $P_0$ the chemical potential of the natural state. 
The stress tensor $\BGs(\mathbf{x},t)=\sigma_{ij}(\mathbf{x},t)\mathbf{e}_i\otimes\mathbf{e}_j$, the heat flux vector $\mathbf{q}(\mathbf{x},t)=q_i(\mathbf{x},t)\mathbf{e}_i$, and the mass flux vector $\mathbf{j}(\mathbf{x},t)=j_i(\mathbf{x},t)\mathbf{e}_i$ are determined, respectively, through the following constitutive relations \citep{Now1,Now2,Now3}
\begin{eqnarray}
\BGs(\mathbf{x},t)&=&\mathfrak{C}\, \bge(\mathbf{x},t)
-
\bga\,
\theta(\mathbf{x},t)
-
\bgb\,
\eta(\mathbf{x},t),\label{eq:Stress}\\
\mathbf{q}(\mathbf{x},t)&=&-
\mathbf{K}\,
\nabla\theta(\mathbf{x},t),\label{eq:ThermalFlux}\\
\mathbf{j}(\mathbf{x},t)&=&-
\mathbf{D}\,
\nabla\eta(\mathbf{x},t),\label{eq:MassFlux}
\end{eqnarray}
with $\bge(\mathbf{x},t)=\textrm{sym}\nabla\bu(\mathbf{x},t)$ denoting the small strains tensor, $\mathfrak{C}=C_{ijkl}\mathbf{e}_i\otimes\mathbf{e}_j\otimes\mathbf{e}_k\otimes\mathbf{e}_l$  the fourth order elasticity tensor showing major and minor symmetries,
$\bga=\alpha_{ij}\mathbf{e}_i\otimes\mathbf{e}_j$  the symmetric second order  thermal dilatation tensor,
$\bgb=\beta_{ij}\mathbf{e}_i\otimes\mathbf{e}_j$   the symmetric second order diffusive expansion tensor,
$\mathbf{K}=K_{ij}\mathbf{e}_i\otimes\mathbf{e}_j$  the symmetric second order  heat conduction tensor, and
$\mathbf{D}=D_{ij}\mathbf{e}_i\otimes\mathbf{e}_j$ the symmetric second order  mass diffusion tensor. 
{
For each constituent layer, the equations of motion,  are given by
\begin{equation}
 \nabla\cdot\BGs(\mathbf{x},t) + \mathbf{b}(\mathbf{x},t)=\rho\,\,\ddot{\bu}(\mathbf{x},t),
\label{Navier}
 \end{equation}
whereas the energy and mass conservation lead, respectively, to the following equations \citep{Now1, Now2, Now3}:
\begin{equation}
 p\dot{\theta}(\mathbf{x},t)+\bga\dot{\bge}(\mathbf{x},t)+\psi\dot{\eta}(\mathbf{x},t)- r (\mathbf{x},t)=-\nabla\cdot\bq(\mathbf{x},t),
\label{Heat}
 \end{equation}
\begin{equation}
 q\dot{\eta}(\mathbf{x},t)+\bgb\dot{\bge}(\mathbf{x},t)+\psi\dot{\theta}(\mathbf{x},t)- s (\mathbf{x},t)=-\nabla\cdot\bj(\mathbf{x},t).
\label{Mass}
 \end{equation}
Term $\rho$ in equation (\ref{Navier}) represents the mass density,  $p$ in equation (\ref{Heat}) is a material constant depending upon the specific heat at constant strain and upon thermodiffusive effects, $q$ in equation (\ref{Mass}) is a material constant related to diffusive effects, and $\psi$ is a material constant measuring thermodiffusive effects \citep{Now1,Now2,Now3}.
Source terms are represented by body forces $\mathbf{b}(\mathbf{x},t)$ in equation (\ref{Navier}), heat sources $r(\mathbf{x},t)$ in equation (\ref{Heat}), and mass sources $s(\mathbf{x},t)$ in equation (\ref{Mass}).
Substituting expressions (\ref{eq:Stress}), (\ref{eq:ThermalFlux}) and (\ref{eq:MassFlux}) into equations \eq{Navier}-\eq{Mass}, one obtains
\begin{eqnarray}
&&\nabla\cdot
\left(
\mathfrak{C}\,\nabla\mathbf{u}(\mathbf{x},t)
\right)
-
\nabla\cdot
\left(\bga\,\theta(\mathbf{x},t)\
\right)
-
\nabla\cdot
\left(\bgb\,\eta(\mathbf{x},t)\
\right)+ \mathbf{b}(\mathbf{x},t) = \rho\,\ddot{\mathbf{u}}(\mathbf{x},t),\label{eq:FieldEquStress}\\
&&\nabla\cdot
\left(\mathbf{K}\,\nabla\theta(\mathbf{x},t)
\right)
-
\bga\,\nabla\dot{\bu}(\mathbf{x},t)
-
\psi\,\dot{\eta}(\mathbf{x},t)
+
r
(\mathbf{x},t)
=
p\,\dot{\theta}(\mathbf{x},t),
\label{eq:FieldEquThermalFlux}\\
&&\nabla\cdot
\left(
\mathbf{D}\,\nabla\eta(\mathbf{x},t)
\right)
-
\bgb\,\nabla\dot{\bu}(\mathbf{x},t)
-
\psi\,\dot{\theta}(\mathbf{x},t)
+
s
(\mathbf{x},t)
=
q\,\dot{\eta}(\mathbf{x},t).\label{eq:FieldEquMassFlux}
\end{eqnarray}
Equations (\ref{eq:FieldEquStress})-(\ref{eq:FieldEquMassFlux}) written in components read
\begin{eqnarray}
&&\left( C_{ijhk} u_{h,k}\right)_{,j}
-
\left(\alpha_{ij} \theta\right)_{,j}
-
\left(\beta_{ij} \eta\right)_{,j}
+
b_i
=
\rho
\ddot{u}_i,
\label{eq:FieldEqStressComponents}
\\
&&\left(K_{ij} \theta_{,i}\right)_{,j}-\alpha_{ij} \dot{u}_{i,j}-\psi\dot{\eta}+r=p\dot{\theta},
\label{eq:FieldEqTemperatureComponents}
\\
&&\left(D_{ij} \eta_{,i}\right)_{,j}-\beta_{ij} \dot{u}_{i,j}-\psi\dot{\theta}+s=q\dot{\eta},
\label{eq:FieldEqChemPotComponents}
\end{eqnarray}
where  $i,j,h,k=1,2$ and  subscript $_,$ denotes the generalized derivative with respect to a spatial coordinate.
 
%
\subsection{Damped Bloch wave propagation in a layered thermo-diffusive material}
\label{SubSec::BlochDecomposition}
According to Floquet-Bloch theory, here generalized for an elastic thermo-diffusive medium, solution of field equations (\ref{eq:FieldEqStressComponents})-(\ref{eq:FieldEqChemPotComponents}) in a periodic laminate material as the one sketched in figure \ref{thermolam}, can be written resorting a Floquet-Bloch like decomposition   in the following way
\begin{equation}
 \mathbf{v}(x_1,x_2,t)=(u_1\,\,  u_2\,\,  \theta\,\,  \eta)^{T}=\mathbf{w}(x_2)\,\textrm{exp}\left[i\left(\mathbf{k}\cdot\mathbf{x}-\omega t\right)\right],
\label{displacement}
 \end{equation}
 where $i$ is the imaginary unit such that $i^2=-1$, $\mathbf{k}=k_1\mathbf{e}_1+k_2\mathbf{e}_2$ is the wave vector, and $\omega$ is the angular frequency.
 In equation (\ref{displacement}), vector $\mathbf{w}(x_2)$ contains the $\mathcal{A}$-periodic Bloch amplitudes of the displacement, temperature, and chemical potential, namely
\begin{equation}
 \mathbf{w}(x_2)=\left(\tilde{u}_1(x_2)\,\, \tilde{u}_2(x_2)\,\, \tilde{\theta}(x_2)\,\, \tilde{\eta}(x_2)\right)^{T}.
\label{amplitude1}
 \end{equation}
 It depends upon the direction of material layering.
 It is worth noting that Floquet-Bloch decomposition (\ref{displacement}) structurally satisfies Floquet-Bloch boundary conditions over the periodic cell $\mathcal{A}$.
 With the aim of investigating free waves propagation inside the laminate, source terms in equations (\ref{eq:FieldEqStressComponents})-(\ref{eq:FieldEqChemPotComponents}) are put to zero ($\mathbf{b}=\mathbf{0}, r=0, s=0$).
 Inserting equation (\ref{displacement}) into field equations (\ref{eq:FieldEqStressComponents})-(\ref{eq:FieldEqChemPotComponents}), by simple algebra, one obtains the following system of partial differential equations expressed in terms of Bloch amplitudes components as dependent variables  and angular frequency $\omega$ and wave vector components as parameters  
\begin{eqnarray}
&&
\left(C_{i2h2}\tilde{u}_{h,2}\right)_{,2}
+
i k_j 
\left[
\left(
C_{ijh2}+C_{i2hj}
\right)
\tilde{u}_{h,2}
+
C_{i2hj,2}\tilde{u}_h
\right]
-
\left(
C_{ijhk} k_k k_j
-
\rho\omega^2\delta_{ih}
\right)
\tilde{u}_h+
\nonumber\\
&&-
\left(\alpha_{i2}\tilde{\theta}
\right)_{,2}
-
i
\alpha_{ij}
\tilde{\theta}
k_j
-
\left(\beta_{i2}\tilde{\eta}
\right)_{,2}
-
i
\beta_{ij}
\tilde{\eta}
k_j=0,
\label{eq:PDE1}
\\
&&
\left(
K_{22}
\tilde{\theta}_{,2}
\right)_{,2}
+
2 i 
K_{2j}
\tilde{\theta}_{,2}
k_j
+
i
K_{i2,2}
\tilde{\theta}
k_i
-
K_{ij}\tilde{\theta} k_i k_j
+
i
\alpha_{i2}
\tilde{u}_{i,2}
 \omega
-
\alpha_{ij}
\tilde{u}_i
k_j \omega
-
i
\psi
\tilde{\eta}
\omega
+
i
p
\tilde{\theta}
\omega
=0,
\label{eq:PDE2}
\\
&&
\left(
D_{22}
\tilde{\eta}_{,2}
\right)_{,2}
+
2 i 
D_{2j}
\tilde{\eta}_{,2}
k_j
+
i
D_{i2,2}
\tilde{\eta}
k_i
-
D_{ij}\tilde{\eta} k_i k_j
+
i
\beta_{i2}
\tilde{u}_{i,2}
 \omega
-
\beta_{ij}
\tilde{u}_i
k_j \omega
-
i
\psi
\tilde{\theta}
\omega
+
i
q
\tilde{\eta}
\omega
=0.
\label{eq:PDE3}
\end{eqnarray}
Generalized derivatives with respect to the $x_1$ coordinate obviously vanish in equations (\ref{eq:PDE1})-(\ref{eq:PDE3}), since layers are stacked along the $x_2$ direction in the considered laminate. 
At this point,in order to investigate propagation and damping of harmonic oscillations in periodic thermodiffusive laminates, it is convenient to determine the transfer matrix for the single homogeneous layer. 
To this aim, partial differential equations (\ref{eq:PDE1})-(\ref{eq:PDE3}) are written in the followings over the single homogeneous layer, thus obtaining a system of second order ordinary  differential equations in the $x_2$-variable.
Once the transfer matrix of a single layer is obtained, by imposing a continuity condition on generalized displacement and traction fields between two adjacent boundaries of two subsequent layers, the transfer matrix of the entire periodic cell $\mathcal{A}$  can be achieved. 
Then, Floquet-Bloch boundary conditions enforced for the periodic cell allow   obtaining  a standard eigenvalue problem, whose characteristic equation  is the dispersion relation of plane oscillations propagating inside the material.
A method considering fixed real-valued wave vectors and complex-valued angular frequencies (usually called $\omega(\mathbf{k})$ formulation) is exploited to investigate temporal damping for the material at hand, while a procedure contemplating  fixed real-valued  angular frequencies and complex-valued wave vectors ($\mathbf{k}(\omega)$ formulation) characterizes the spatial decay  of waves propagating inside the medium. 
These two formulations are described in Section \ref{Sec:TransferMatrixMethod}, while subsequent illustrative examples are focused on the investigation of spatial damping inside  the periodic laminate with the aim of studying the influence of thermal and diffusive coupling upon the band diagram of mechanical waves travelling inside the material.
A procedure which could be exploited in order to investigate temporal damping is detailed in Appendix D.

\subsection{Field equations for a single layer in terms of Bloch amplitudes  }
\label{SubSec::WaveLikeSolutionForASingleLayer}
Equations (\ref{eq:PDE1})-(\ref{eq:PDE3}) written for the single homogeneous layer of the laminate represented in figure \ref{thermolam} take the form
\begin{eqnarray}
&&
C_{1212}\tilde{u}_{1,22}
+
2 i k_2 C_{1212}\tilde{u}_{1,2}+ i k_1\left( C_{1212}+C_{1122}\right) \tilde{u}_{2,2}
+
\left(\rho\omega^2-k_1^2 C_{1111}-k_2^2 C_{1212}\right)\tilde{u_1}
+
\nonumber\\
&&
-
k_1 k_2\left(C_{1212}+C_{1122}\right)\tilde{u_2}
-
i k_1\alpha_{11}\tilde{\theta}
-
i k_1\beta_{11}\tilde{\eta}=0,
\label{eq:ODE1}
\\
&&
C_{2222}\tilde{u}_{2,22}
+
2 i k_2 C_{2222}\tilde{u}_{2,2}+ i k_1\left( C_{1212}+C_{1122}\right) \tilde{u}_{1,2}
-
\alpha_{22}\tilde{\theta}_{,2}
-
\beta_{22}\tilde{\eta}_{,2}
+
\nonumber\\
&&
+
\left(\rho\omega^2-k_1^2 C_{1212}-k_2^2 C_{2222}\right)\tilde{u_2}
-
k_1 k_2\left(C_{1212}+C_{1122}\right)\tilde{u_1}
-
i k_2\alpha_{22}\tilde{\theta}
-
i k_2\beta_{22}\tilde{\eta}=0,
\label{eq:ODE2}
\\
&&
K_{22}\tilde{\theta}_{,22}
+
i\omega\alpha_{22}\tilde{u}_{2,2}
+
2i k_2 K_{22}\tilde{\theta}_{,2}
-
\omega k_1 \alpha_{11} \tilde{u}_1
-
\omega k_2 \alpha_{22} \tilde{u}_2
+
\left(
 i \omega p
-k_1^2 K_{11}
-
k_2^2 K_{22}
\right)
\tilde{\theta}
+
\nonumber\\
&&+
i\psi\omega\tilde{\eta}=0,
\label{eq:ODE3}
\\
&&
D_{22}\tilde{\eta}_{,22}
+
i\omega\beta_{22}\tilde{u}_{2,2}
+
2i k_2 D_{22}\tilde{\eta}_{,2}
-
\omega k_1 \beta_{11} \tilde{u}_1
-
\omega k_2 \beta_{22} \tilde{u}_2
+
\left(
 i \omega q
-k_1^2 D_{11}
-
k_2^2 D_{22}
\right)
\tilde{\eta}
+
\nonumber\\
&&+
i\psi\omega\tilde{\theta}=0.
\label{eq:ODE4}
\end{eqnarray}
where, this time, derivatives  with respect to the  spatial coordinate $x_2$, are considered as classical derivatives.
Second order ordinary differential equations (\ref{eq:ODE1})-(\ref{eq:ODE4}) can  be written in operatorial form as
\begin{equation}
 \mathbf{A}\,\mathbf{w}^{''}+\mathbf{B}\,\mathbf{w}^{'}+\mathbf{C}\,\mathbf{w}=\mathbf{0}, 
\label{ODEs}
 \end{equation}
where apex $'$ denotes the derivative with respect to the $x_2-$variable, and the  $4\times4$ matrices $\mathbf{A}$, $\mathbf{B}$ and $\mathbf{C}$ are given by
\begin{eqnarray}
&&\mathbf{A}=
\left(
\begin{array}{cccc}
 C_{1212} & 0 & 0 & 0 \\
 0 &  C_{2222} & 0 & 0 \\
 0 & 0 & K_{22} & 0 \\
 0 & 0 & 0 & D_{22}
\end{array}
\right), 
\nonumber\\
&&
\nonumber\\
&& \mathbf{B}= 
\left(
\begin{array}{cccc}
 2ik_{2}C_{1212} & ik_{1}(C_{1212}+C_{1122}) & 0 & 0 \\
  ik_{1}(C_{1122}+C_{1212}) &  2ik_{2}C_{2222} & -\alpha_{22} & -\beta_{22} \\
 0 & i\omega\alpha_{22} & 2ik_{2}K_{22} & 0 \\
 0 & i\omega\beta_{22} & 0 & 2ik_{2}D_{22}
\end{array}
\right),
\label{mAB}
\nonumber\\
&&
\nonumber\\
&& \mathbf{C}= 
\left(
\begin{array}{cccc}
 \left(\begin{array}{c}
 \rho\omega^2\\
 -k_1^2C_{1111}\\
 -k_2^2C_{1212}
 \end{array}\right) & -k_1 k_2 \left(C_{1122}+C_{1212}\right) & -ik_1\alpha_{11} & -ik_1\beta_{11} \\
  -k_1 k_2 \left(C_{1122}+C_{1212}\right) &  \left(\begin{array}{c}
  \rho\omega^2\\
  -k_1^2C_{1212}\\
  -k_2^2C_{2222}
  \end{array}\right) & -ik_2\alpha_{22} & -ik_2\beta_{22} \\
 -\omega k_1 \alpha_{11} & -\omega k_2 \alpha_{22} & \left(\begin{array}{c}
 i\omega p\\
 -k_1^2K_{11}\\
-k_2^2K_{22}
 \end{array}\right) & i\omega\psi \\
 -\omega k_1 \beta_{11} & -\omega k_2 \beta_{22} & i\omega\psi & \left(\begin{array}{c}
 i\omega q\\
 -k_1^2D_{11}\\
 -k_2^2 D_{22}
 \end{array}\right)
\end{array}
\right). \nonumber\\
\label{mC}
\end{eqnarray}
The general formal solution of  system \eq{ODEs} is reported in details in the next Section for the most general case where  thermodiffusive effects are coupled with  mechanical displacement and stresses.
}
%
%
%
\section{Transfer matrix method to determine the frequency band structure of a laminate composite}
\label{Sec:TransferMatrixMethod}
\begin{figure}[h!]
\centering
\includegraphics[scale=0.48,angle=-90]{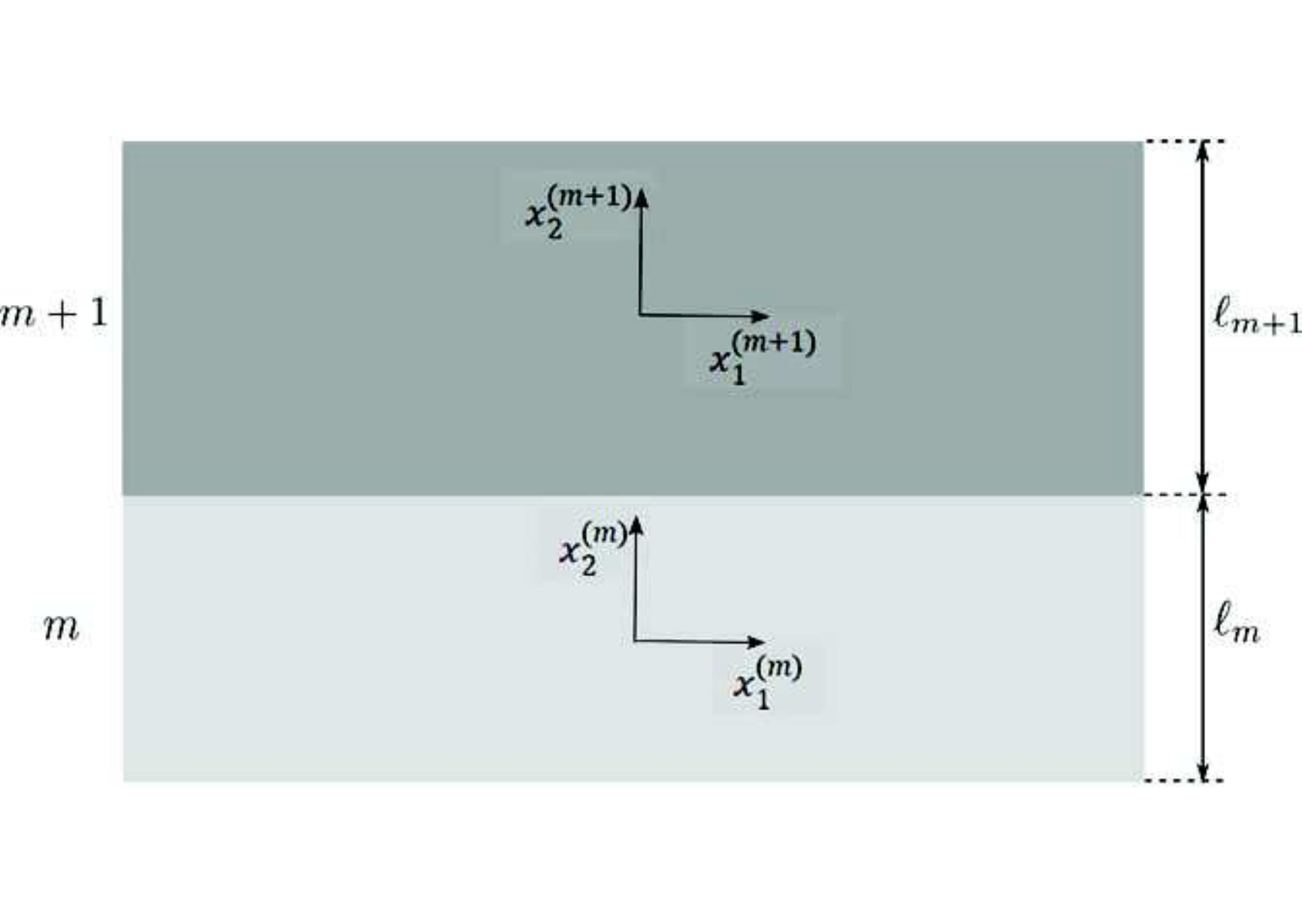}
 \protect\protect\caption{Two subsequent layers of arbitrary thickness belonging to the periodic cell. Local systems of coordinates used for deriving the transfer matrix of each layer are reported.\label{bilayer}}
\end{figure}

{Introducing the eight-components vector $\mathbf{r}=(\mathbf{w}^{'}\,\, \mathbf{w})^{T}$, one can easily transform the second order $4\times4$ system \eq{ODEs} in the following equivalent first order $8\times8$ system
\begin{equation}
\mathbf{M}\mathbf{r}^{'}+\mathbf{N}\mathbf{r}=\mathbf{0},
\label{ODEr}
\end{equation}
where $\mathbf{M}$ is a non singular square diagonal block matrix and $\mathbf{N}$ is a square block matrix. They  are $8\times8$ matrices expressed, respectively, as}
\beq
 \mathbf{M}=
\left(
\begin{array}{cc}
 \mathbf{A} &  \mathbf{0} \\
 \mathbf{0} &  \mathbf{I}
\end{array}
\right),
\quad 
\mathbf{N}=
\left(
\begin{array}{cc}
 \mathbf{B} &  \mathbf{C} \\
 -\mathbf{I} &  \mathbf{0}
\end{array}
\right).
\label{eq:M&N}
\eeq
General solution of first order ordinary differential  system (\ref{ODEr}) can be written as
\begin{equation}
\mathbf{r}=\textrm{exp}\left[\mathbf{M}^{-1}\mathbf{N} x_2\right]\,\mathbf{c},
\label{eq:solutionr}
\end{equation}
where  $\mathbf{c}$ is a vector of constants and  $\textrm{exp}[\,\cdot\,]$ denotes the matrix exponential.
A possible procedure to compute matrix exponential is detailed in Appendix A.
{
Denoting with $\mathbf{y}(x_1,x_2,t)$ a vector containing the components of solution vector $\mathbf{v}(x_1,x_2,t)$ of equation (\ref{displacement}) and of the generalized traction vector $\mathbf{s}(x_1,x_2,t)$ defined as
\begin{equation}
 \mathbf{s}(x_1,x_2,t)=(\sigma_{21}\,\, \sigma_{22}\,\, q_{2}\,\,j_{2})^{T}=\mathbf{t}(x_2)\,\textrm{exp}\left[i\left(\mathbf{k}\cdot\mathbf{x}-\omega t\right)\right],
\label{traction}
 \end{equation}
 where $\mathbf{t}$ is given by
\begin{equation}
 \mathbf{t}(x_2)=\left(\tilde{\sigma}_{21}(x_2)\,\, \ \tilde{\sigma}_{22}(x_2)\,\, \ \tilde{q}_{2}(x_2)\,\, \ \tilde{j}_{2}(x_2)\right)^{T},
\label{amplitude2}
 \end{equation}
it can be expressed in terms of $\mathbf{r}$ in the following way
\begin{equation}
\mathbf{y}(x_1,x_2,t)
=
\left(\begin{array}{c}
\mathbf{v}(x_1,x_2,t)\\
\mathbf{s}(x_1,x_2,t)
\end{array}\right)
=
\textrm{exp}\left[i(\mathbf{k}\cdot\mathbf{x}-\omega t)\right]
\left(
\begin{array}{c c}
\mathbf{0}&\mathbf{I}\\
\mathbf{R}&i\mathbf{R}k_2+\mathbf{S}
\end{array}
\right)\,\mathbf{r},
\label{eq:vectory}
\end{equation}
where $\mathbf{I}$ is a $4\times4$ identity operator and non singular diagonal matrix $\mathbf{R}$ and coupling singular matrix $\mathbf{S}$ are expressed, respectively, as
\begin{equation}
\mathbf{R}
=
\left(
\begin{array}{c c c c }
C_{1212} & 0 & 0 & 0\\
0 & C_{2222} & 0 & 0\\
0 & 0 & -K_{22} & 0\\
0 & 0 & 0 & -D_{22}
\end{array}
\right),
\hspace{0.5cm}
\mathbf{S}
=
\left(
\begin{array}{c c c c }
0 & 0 & 0 & 0\\
0 & 0 & -\alpha_{22} & -\beta_{22}\\
0 & 0 & 0 & 0\\
0 & 0 & 0 & 0
\end{array}
\right).
\end{equation}
}
Plugging solution (\ref{eq:solutionr}) into (\ref{eq:vectory}) one obtains
\begin{equation}
\mathbf{y}(x_1,x_2,t)=
\left(
\begin{array}{c c}
\mathbf{0}&\mathbf{I}\\
\mathbf{R}&i\mathbf{R}k_2+\mathbf{S}
\end{array}
\right)
\textrm{exp}\left[-\mathbf{M}^{-1}\mathbf{N}x_2\right]\mathbf{c}\,\textrm{exp}\left[i(\mathbf{k}\cdot\mathbf{x}-\omega t)\right].
\end{equation}
If the single $m^{th}$ layer belonging to the periodic cell shown in figure \ref{thermolam} has thickness $\ell_{m}$, referring to a local coordinate system as the one depicted in figure \ref{bilayer}, such as along the $x_2-$axis the layer extends in the range $-\ell_m/2\leq x_2\leq\ell_m/2$,  one can define the generalized vector $\mathbf{y}$ containing displacement components, relative temperature, relative chemical potential, tractions, heat and mass fluxes at the upper and lower boundaries of the layer as
\begin{eqnarray}
\mathbf{y}_m^+&=&\mathbf{y}_m(x_1,x_2=\ell_m/2,t)=
\left(
\begin{array}{c c}
\mathbf{0} & \mathbf{I}\\
\mathbf{R} & i\mathbf{R}k_2+\mathbf{S}
\end{array}
\right)
\textrm{exp}\left[-\mathbf{M}^{-1}\mathbf{N}\ell_m/2\right]\mathbf{c}\,\,\textrm{exp}\left[i(k_1x_1+k_2\ell_m/2-\omega t)\right],
\label{eq:y+}\nonumber\\
\\
\mathbf{y}_m^-&=&\mathbf{y}_m(x_1,x_2=-\ell_m/2,t)=
\left(
\begin{array}{c c}
\mathbf{0} & \mathbf{I}\\
\mathbf{R} & i\mathbf{R}k_2+\mathbf{S}
\end{array}
\right)
\textrm{exp}\left[\mathbf{M}^{-1}\mathbf{N}\ell_m/2\right]\mathbf{c}\,\,\textrm{exp}\left[i(k_1x_1-k_2\ell_m/2-\omega t)\right].
\nonumber\\
\label{eq:y-}
\end{eqnarray}
{Since block matrix premultiplying the exponential matrix is non singular by definition, from equation (\ref{eq:y-}) constants vector $\mathbf{c}$ gains the form}
\begin{equation}
\mathbf{c}=\textrm{exp}\left[-\mathbf{M}^{-1}\mathbf{N}\ell_m/2\right]
\left(
\begin{array}{c c}
\mathbf{0} & \mathbf{I}\\
\mathbf{R} & i\mathbf{R}k_2+\mathbf{S}
\end{array}
\right)^{-1}
\mathbf{y}^-_m\,\textrm{exp}\left[i(k_1x_1-k_2\ell_m/2-\omega t)\right].
\label{eq:vectorconstantsc}
\end{equation}
Substitution of expression (\ref{eq:vectorconstantsc}) into (\ref{eq:y+}) leads to express $\mathbf{y}_m^+$ in terms of $\mathbf{y}_m^-$ as
\begin{equation}
\mathbf{y}_m^+=\left(
\begin{array}{c c}
\mathbf{0} & \mathbf{I}\\
\mathbf{R} & i\mathbf{R}k_2+\mathbf{S}
\end{array}
\right)
\textrm{exp}\left[-\mathbf{M}^{-1}\mathbf{N}\ell_m\right]
\left(
\begin{array}{c c}
\mathbf{0} & \mathbf{I}\\
\mathbf{R} & i\mathbf{R}k_2+\mathbf{S}
\end{array}
\right)^{-1}
\,\textrm{exp}\left[{ik_2\ell_m}\right]
\mathbf{y}_m^-
=
\mathbf{T_m}\,\mathbf{y}_m^-,
\label{eq:y+New}
\end{equation}
where $\mathbf{T}_m$ is the frequency-dependent transfer matrix of the $m^{th}$ thermodiffusive elastic layer \citep{Gupta1970,Faulkner1985}. 
Since relation (\ref{eq:y+New}) is valid for each single layer forming the periodic cell and since 
it is assumed that the layers are perfectly bonded, so that continuity condition 
\beq
\mathbf{y}^{+}_{m}=\mathbf{y}^{-}_{m+1}
\label{continuity}
\eeq
must be satisfied at the interface between two subsequent layers $m$ and $m+1$ (see figure \ref{bilayer}),
the following equation can be easily derived relating generalized vector at the upper boundary of the last $n^{th}$ layer $\mathbf{y}_n^+$ to the generalized vector at the lower boundary of the first layer $\mathbf{y}_1^-$. It reads
\begin{equation}
\mathbf{y}_n^+=\mathbf{T}_{(1,n)}\,\mathbf{y}_1^-,
\label{eq:yn+}
\end{equation}
where $\mathbf{T}_{(1,n)}=\prod_{i=0}^{n-1}\mathbf{T}_{n-i}$ is the frequency-dependent transfer matrix of the entire periodic cell.

\begin{figure}[h!]
  \centering
  \begin{tabular}{c c }
 \hspace{-0.5cm}
  \includegraphics[width=6cm,angle=-90]{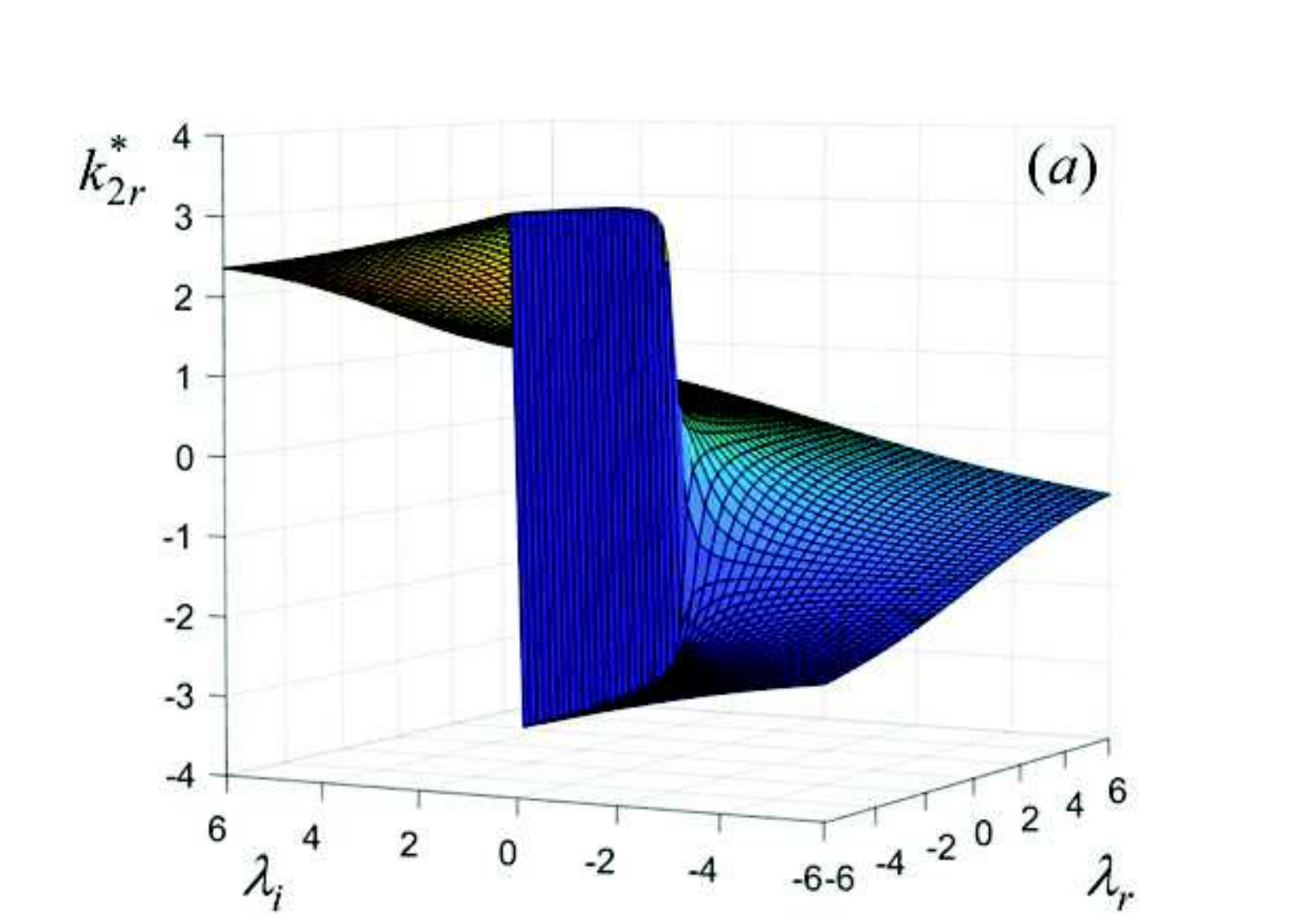}
  &
 \hspace{-0.5cm}
  \includegraphics[width=6cm,angle=-90]{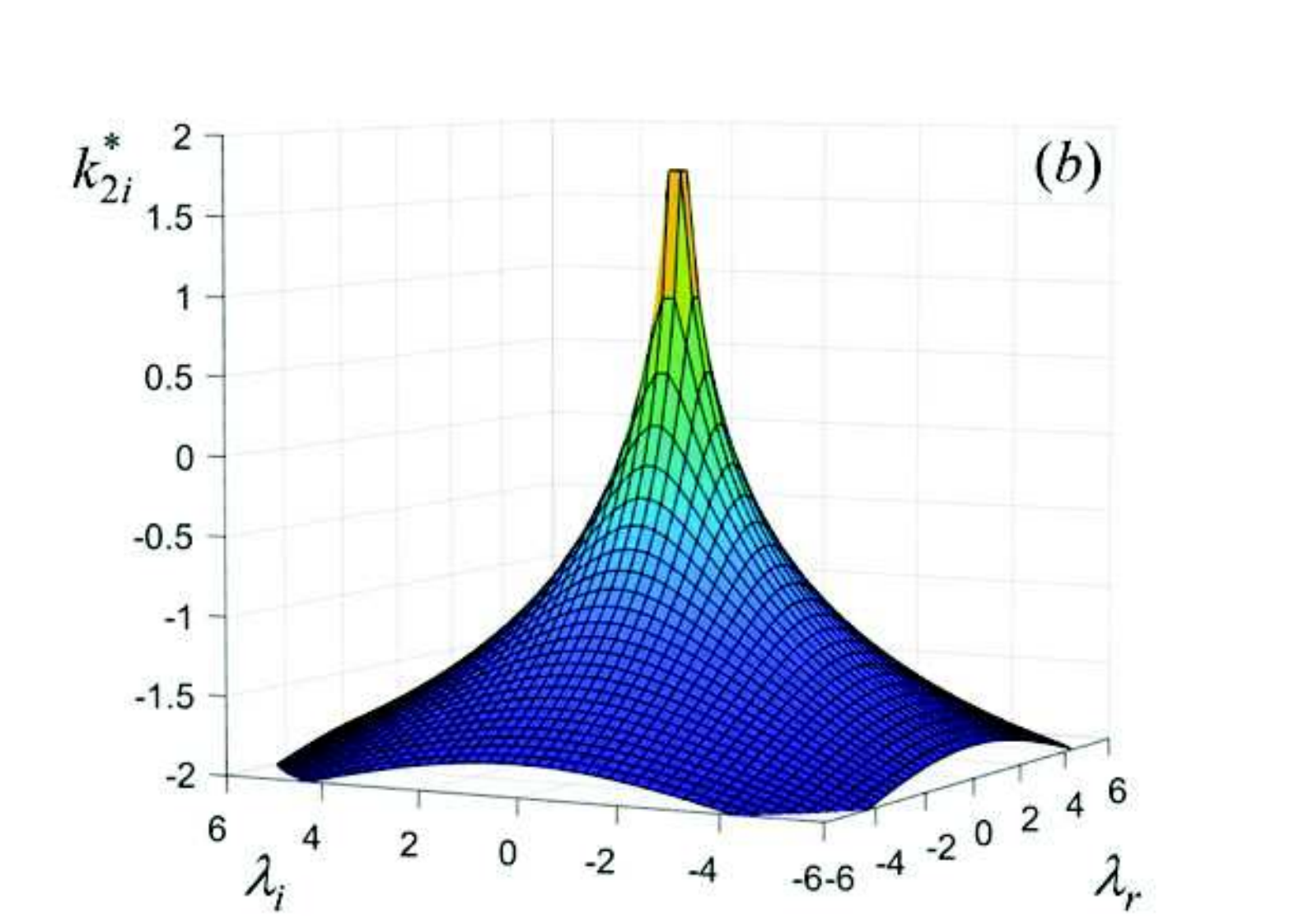}
  \end{tabular}
  \caption{\it (a) dimensionless wave number  $k_{2r} ^*$ as a function of the real and imaginary parts of the Floquet multiplier $\lambda$. (b) dimensionless wave number  $k_{2i} ^*$ as a function of the real and imaginary parts of the Floquet multiplier $\lambda$ }
  \label{Fig::kappa2_lambda}
\end{figure}

{In virtue of the periodicity of cell $\mathcal{A}$, 
the following  Floquet-Bloch boundary condition \citep{Floquet1883,Bloch1929,Brillouin1953,Mead1973,Langley1993} can be imposed
\beq
\mathbf{y}^{+}_{n}=
\textrm{exp}\left[ik_2L\right]
\mathbf{y}^{-}_{1},
\label{FB}
\eeq
where, due to the geometry of the system (see figure \ref{thermolam}), the periodicity direction is assumed to be along the $x_{2}-$axis and, as already mentioned, $L=\sum_{m=1}^n\ell_m$ is the extent of the whole periodic cell along that direction.}
Substituting  (\ref{FB}) into (\ref{eq:yn+}), one obtains the following standard eigenvalue problem
\begin{equation}
\left(
\mathbf{T}_{(1,n)}-\lambda\mathbf{I}
\right)
\mathbf{y}_1^-=\mathbf{0},
\label{eq:FBsyst}
\end{equation}
where  $\lambda=\textrm{exp}\left[ik_2L\right]$ is called Floquet multiplier and $\mathbf{I}$ represents an $8\times 8 $ identity operator.
The system (\ref{eq:FBsyst}) admits a non-trivial solution when the following characteristic equation is satisfied
\beq
\mD(\mathbf{k},\omega)=\textrm{Det}\left(\mathbf{T}_{(1,n)}-\lambda\mathbf{I}\right)=0.
\label{FBdisp}
\eeq
Equation \eq{FBdisp} is the dispersion relation of plane oscillations in periodic thermodiffusive laminates where the elementary cell is composed by an 
arbitrary number of layers $n$. %
Furthermore,
transfer matrix $\mathbf{T}_{(1,n)}$ results to be a symplectic matrix having a unitary determinant. 
In the most general case, both the wave vector $\mathbf{k}$ and the angular frequency $\omega$, to which characteristic equation $\mathcal{D}$ depends, can be complex, namely $\mathbf{k}=(k_{1r}+i\, k_{1i})\mathbf{e}_1+(k_{2r}+i\, k_{2i})\mathbf{e}_2$ and $\omega=\omega_r+i\,\omega_i$.
In this case, wave vector $\mathbf{k}$ can be specialized in the form
\begin{equation}
\mathbf{k}=\mathbf{k}_r+ i \mathbf{k_i}=k_r\,\mathbf{n}_r+i\,k_i\,\mathbf{n}_i,
\label{eq:wavevector}
\end{equation}
where $\mathbf{k}_r$ represents the real wave vector having magnitude $k_r$ and direction $\mathbf{n}_r\in\mathbb{R}^2$, and $\mathbf{k}_i$ is the attenuation vector with magnitude $k_i$ and direction $\mathbf{n}_i\in\mathbb{R}^2$.
A plane wave can be defined as homogeneous when the direction of normals to planes of constant phase $\mathbf{n}_r$ coincides with the one of normals to planes of constant amplitude $\mathbf{n}_i$, namely when $\mathbf{n}_r\times\mathbf{n}_i=\mathbf{0}$ \citep{Carcione2007}.
Denoting with $\mathbf{n}$ such a direction one has
\begin{equation}
\mathbf{k}=\left(k_r+i\,k_i\right)\mathbf{n}=\kappa\, \mathbf{n},
\label{eq:wavevector2}
\end{equation}
with $\kappa$ the complex wave number. 
Furthermore, being $\mathbf{k}_r/k_r=\mathbf{k}_i/k_i$, for an homogeneous wave one obtains the following relation among the real and imaginary parts of $k_1$ and $k_2$
\begin{equation}
k_{1r}\,k_{2i}=k_{2r}\,k_{1i}.
\label{eq:componentsrelationhomogeneouswave}
\end{equation}
When $\mathbf{k}\in\mathbb{C}^2$ and $\omega\in\mathbb{C}$,  frequency spectrum is determined from the intersection of two hypersurfaces immersed in a space in $\mathbb{R}^6$, representing respectively the vanishing of the real and imaginary part of characteristic equation (\ref{FBdisp}), namely
\begin{equation}
\left\{
\begin{array}{l}
\textrm{Re}\left( \mathcal{D}\left( k_{1r},k_{1i},k_{2r},k_{2i},\omega_{r},\omega_{i}\right))\right)=0\\
\textrm{Im}\left(\mathcal{D}\left(k_{1r},k_{1i},k_{2r},k_{2i},\omega_{r},\omega_{i}\right)\right)=0
\end{array}
\right.
.
\label{eq:Hypersurfaces}
\end{equation}
In order to investigate spatial damping for the material at hand, the wave vector $\mathbf{k}$  is considered  as complex ($k_{\alpha}=k_{\alpha r}+i\,k_{\alpha i}$ with $\alpha=1,2$) and the angular frequency $\omega$ as real \citep{Caviglia1992}.
In the particular case where the value of one component $k_\alpha$  is fixed ($\alpha=1$ or $\alpha=2$), frequency spectrum is obtained through the intersection of two surfaces in $\mathbb{R}^3$, namely the plane $\{k_{\beta r},k_{\beta i},\omega\}$, with $\beta\neq\alpha$, as
\begin{equation}
\left\{
\begin{array}{l}
\textrm{Re}(\mD(k_{\beta r},k_{\beta i},\omega))=0\\
\textrm{Im}(\mD(k_{\beta r},k_{\beta i},\omega))=0\\
\end{array}
\right.
,
\label{eq::kcomplex_omegareal}
\end{equation}
and, if fixing $k_\alpha$ equation (\ref{eq:componentsrelationhomogeneouswave}) results satisfied, the plane wave is homogeneous.
By fixing, for example, component $k_1$ of complex wave vector $\mathbf{k}$, 
a procedure for obtaining material frequency band structure that is alternative to (\ref{eq::kcomplex_omegareal}) is to directly solve linear eigenvalue problem (\ref{eq:FBsyst}), where the Floquet multiplier $\lambda$ is the eigenvalue and $\mathbf{y}_1^-$ is the eigenvector.
 In this  situation, in fact, it is possible to prove that transfer matrix $\mathbf{T}_{(1,n)}$ results to be independent upon $k_2$ and characteristic equation (\ref{FBdisp}) reduces to the $8^{th}$-degree associated polynomial.
In this case, being the wave number related to the Floquet multiplier by relation $k_2= \textrm{ln}(\lambda)/(i\,L)$, its real and imaginary parts are expressed in terms of $\lambda=\lambda_r+i\,\lambda_i$ as
\begin{equation}
k_{2r}=\frac{\textrm{Arg}(\lambda_r+i\,\lambda_i)}{L},
\hspace{0.5cm}
k_{2i}=-\frac{1}{2}\frac{\textrm{ln}(\lambda_r^2+\lambda_i^2)}{L},
\label{eq:wavenumberk2}
\end{equation}
where symbol $\textrm{Arg}(\cdot)$ denotes the argument of a complex number.
As expected, $k_{2r}L$ is a  function whose values belong to the first, dimensionless, Brillouin zone $(-\pi,\pi]$. Figure \ref{Fig::kappa2_lambda} shows the behaviour of dimensionless wave numbers $k_{2r} ^*=k_{2r} L$ and $k_{2i}^*=k_{2i} L$ in terms of the real and imaginary parts of Floquet multiplier $\lambda$. As depicted in figure \ref{Fig::kappa2_lambda}-(a), $k_{2r}^*$ shows a branch cut discontinuity in the complex $\lambda$ plane running from $-\infty$ to $0$.
Moreover, since $\mathbf{T}_{(1,n)}$  is a symplectic matrix, if $\lambda_k$ is the $k^{th}$ eigenvalue for characteristic equation (\ref{FBdisp}), also $1/\lambda_k$ is an eigenvalue.
Such eigenvalues, in fact, are the roots of a palindromic characteristic polynomial, which 
is characterized by a reduced number of invariants \citep{Hennig1999,Romeo2002,Bronski2005,Xiao2013,Carta2015,Carta2016}.
A procedure to compute the invariants of such characteristic polynomial is detailed in Appendix B.
When component $k_2$ of $\mathbf{k}$ is fixed, in order to study wave  propagation in the $\mathbf{e}_1$ direction, one could exploit the formal solution outlined in  Appendix C, which 
 allows expressing transfer matrix $\mathbf{T}_m$ of the single $m^{th}$ layer as a power series of wave number $k_1$.
In this way, by combining transfer matrices of all the $n$ layers constituting the periodic cell, one  obtains the transfer matrix of the entire cell $\mathbf{T}_{(1,n)}$ as a power series of $k_1$.
Truncating this last at a proper order, it is possible to obtain an approximation of the eigenvalue problem (\ref{FBdisp}) showing a polynomial dependence upon $k_1$, which can be used to investigate propagation of plane waves in the $\mathbf{e}_1$ direction.
Temporal damping is studied by considering the angular frequency $\omega$ in \eqref{FBdisp}  as complex ($\omega=\omega_r+i\,\omega_i$) and wave vector $\mathbf{k}$ as real \citep{Carcione2007}.
In this case, once a component of $\mathbf{k}$ is fixed ($k_\alpha$ with $\alpha=1$ or $2$), frequency spectrum is obtained by means of the intersection between two surfaces in $\mathbb{R}^3$, namely  the plane $\{k_\beta,\omega_r,\omega_i\}$, with $\beta\neq\alpha$. Such surfaces represent  the vanishing of the real and imaginary parts of implicit function $\mD$, namely 
\begin{equation}
\left\{
\begin{array}{l}
\textrm{Re}(\mD(k_\beta,\omega_r,\omega_i))=0\\
\textrm{Im}(\mD(k_\beta,\omega_r,\omega_i))=0\\
\end{array}
\right.
.
\end{equation}
Analogously to what done for spatial damping, Appendix D describes a formal procedure to express transfer matrix of a single layer as a power series of angular frequency $\omega$. Following the same path of reasoning as before, transfer matrix of the entire periodic cell can thus be truncated at a proper order of $\omega$ in order to obtain a useful approximation of the eigenvalue problem (\ref{FBdisp}) with a polynomial dependence upon $\omega$ with the aim of investigating temporal damping for the material at hand.
%
\section{Illustrative examples}
\label{Sec::Numerical results}
%
Solution of the general characteristic equation \eq{FBdisp} is  performed in the followings for  thermodiffusive multi-layered systems of interest for
engineering and technology applications. 
In particular, the behaviour of  a thermodiffusive bi-layered composite which can be used in the fabrication of solid oxide fuel cells (SOFCs) \citep{Bacigalupo7, Bacigalupo9, fantoni2020}, is explored.
{Focusing the attention upon spatial damping inside the system, the linear eigenvalue problem (\ref{eq:FBsyst}) has been solved in terms of the Floquet multiplier $\lambda$.} 
Referring to coordinate system represented in figure \ref{bilayer}, for a fixed value of $k_1$, the behaviour of  real and imaginary parts of $k_2$, related, respectively, to the propagating part and to the spatial attenuation of the wave, is investigated with respect to the real independent parameter $\omega$.
By means of a parametric analysis,  the effects of the coupling between thermal, diffusive and mechanical fields on the 
dispersion and damping curves as well as their physical implications are discussed in details.

\subsection{Dispersion and damping in bi-phase thermodiffusive layered media of interest for SOFC devices fabrication}
\label{SOFCex}
One considers a periodic bi-phase laminate composed by materials of interest for solid oxide fuel cells fabrication, similar to those introduced in 
\citealt{Bacigalupo8}.
Phase 1, representing the SOFC's ceramic electrolyte, is  assumed to be constituted by Yttria-stabilized zirconia (YSZ), whereas 
 phase 2, representing an electrode (cathode or anode), is assumed to be made by a Nichel-based ceramic-metallic composite material (see for example 
\citealt{Zhu1}, \citealt{Brandon1}). 
Propagation of plane harmonic Bloch waves which can be modelled using expression \eq{displacement}, is explored. 
In the calculations, both layers are considered to  have the same thickness $\ell_1=\ell_2=1 \ \textrm{mm}$.
Assuming a plane strain condition and isotropic phases constitutive equations (\ref{eq:Stress})-(\ref{eq:MassFlux}) simplifies into
\begin{eqnarray}
&&\BGs(\mathbf{x},t)=2G\bge(\mathbf{x},t)+
\left(\frac{2\nu G}{1-2\nu}\trace{\bge(\mathbf{x},t)}-\alpha\theta(\mathbf{x},t)-\beta\eta(\mathbf{x},t)\right)\mathbf{I},\label{eq:StressIsotr}\\
&&\mathbf{q}(\mathbf{x},t)=-K\nabla\theta(\mathbf{x},t),\label{eq:ThermaFluxIsotr}\\
&&\mathbf{j}(\mathbf{x},t)=-D\nabla\eta(\mathbf{x},t)\label{eq:MassFluxIsotr},
\end{eqnarray}
with shear modulus $G$ expressed in terms of Young's modulus $E$ and Poisson ration $\nu$ as $G=E/(2(1+\nu))$,  $\alpha={2G(1+\nu)\alpha_t}/{(1-2\nu)}$ being $\alpha_t$ the coefficient of linear thermal dilation,  $\beta={2G(1+\nu)\beta_t}/(1-$ $2\nu)$ being $\beta_t$ the coefficient of linear diffusion dilation, thermal conductivity constant $K$, and mass diffusivity constant $D$.
For the phase 1 (YSZ-electrolyte), 
the values of the Young's modulus, Poisson's ratio and mass density are assumed to be, respectively, $E_{1}=155 \ \textrm{GPa}$, $\nu_{1}=0.3$ and 
$\rho_{1}=5532 \ \textrm{kg}/\textrm{m}^{3}$, whereas for the phase 2 (Ni-based composite) they are $E_{2}=50 \ \textrm{GPa}$, $\nu_{2}=0.25$ and 
$\rho_{2}=6670 \ \textrm{kg}/\textrm{m}^{3}$ (see \citealt{Johnson1}, \citealt {Anand1} and  \citealt{Nakajo1}).
Concerning the thermal properties of the layers, the thermal 
conductivities of the phases are $K_{t1}=2.64 \ \textrm{W}/\textrm{mK}$ and  $K_{t2}=9.96 \ \textrm{W}/\textrm{mK}$, the specific heats 
$C_1=400 \ \textrm{J}/\textrm{kgK}$ and $C_2=440 \ \textrm{J}/\textrm{kgK}$ and the temperature of the natural state is assumed to be 
$T_{0}=293.15\, \textrm{K}$.
 The normalized thermal conductivity and the thermodiffusive coefficient $p_i$  introduced in the governing equations
(\ref{eq:FieldEquStress})-(\ref{eq:FieldEquMassFlux}) are  given, respectively, by $K_{i}=K_{ti}/T_{0}$ and $p_{i}=\rho_{i}C_{i}/T_{0}, \ \ i=1,2$. 
Coefficients of linear thermal dilatation are given by $\alpha_{t1}=2.2205\cdot 10^{-6}\,1/K$ and $\alpha_{t2}=3.8858\cdot 10^{-6}\,1/K$, while coefficients of  linear diffusion dilatation $\beta_{ti}$ $(i=1,2)$ are assumed to have a value equal to $1/10$ of the correspondent $\alpha_{ti}$.
Regarding the diffusive properties of the 
two layers, the ratio between the diffusion coefficient $D_i$ and the thermodiffusive coefficient $q_i$ used in equation 
(\ref{eq:FieldEquMassFlux}) are assumed to be equal to $D_1/q_1=0.9\cdot 10^{-5} \textrm{m}^{2}/\textrm{s}$ and $D_2/q_2=0.73\cdot 10^{-5} \textrm{m}^{2}/\textrm{s}$, with the value of $q_i$ equal to $1/10$ of the respective $p_i$ $(i=1,2)$. 
Finally, thermodiffusive coupling coefficients $\psi_i$ are taken with a value equal to $1/3$ of the correspondent $p_i$.

 \begin{figure}
  \centering
  \begin{tabular}{c c }
 \hspace{-0.75cm}
  \includegraphics[width=6cm,angle=-90]{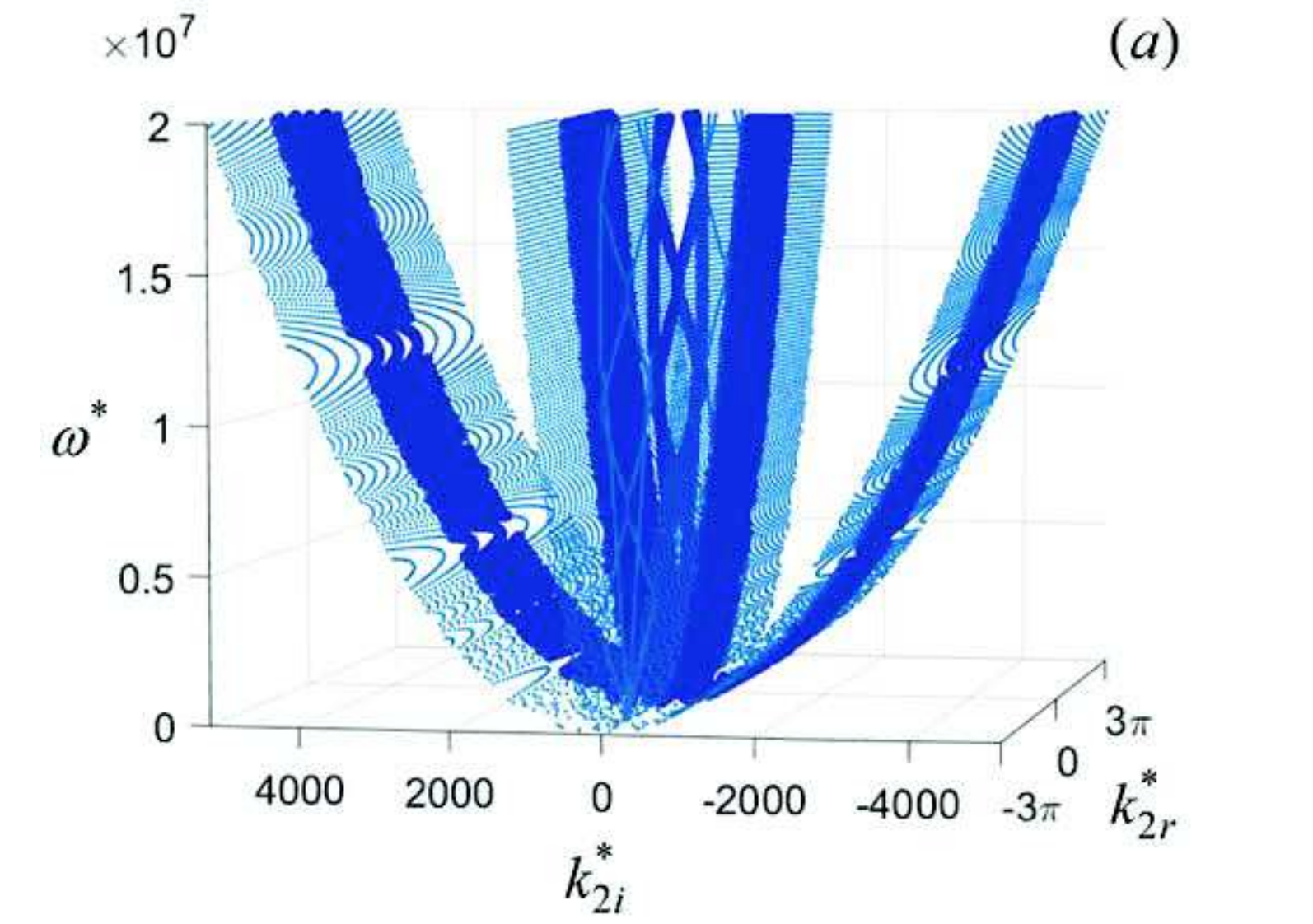}
  &
 \hspace{-1cm}
  \includegraphics[width=6cm,angle=-90]{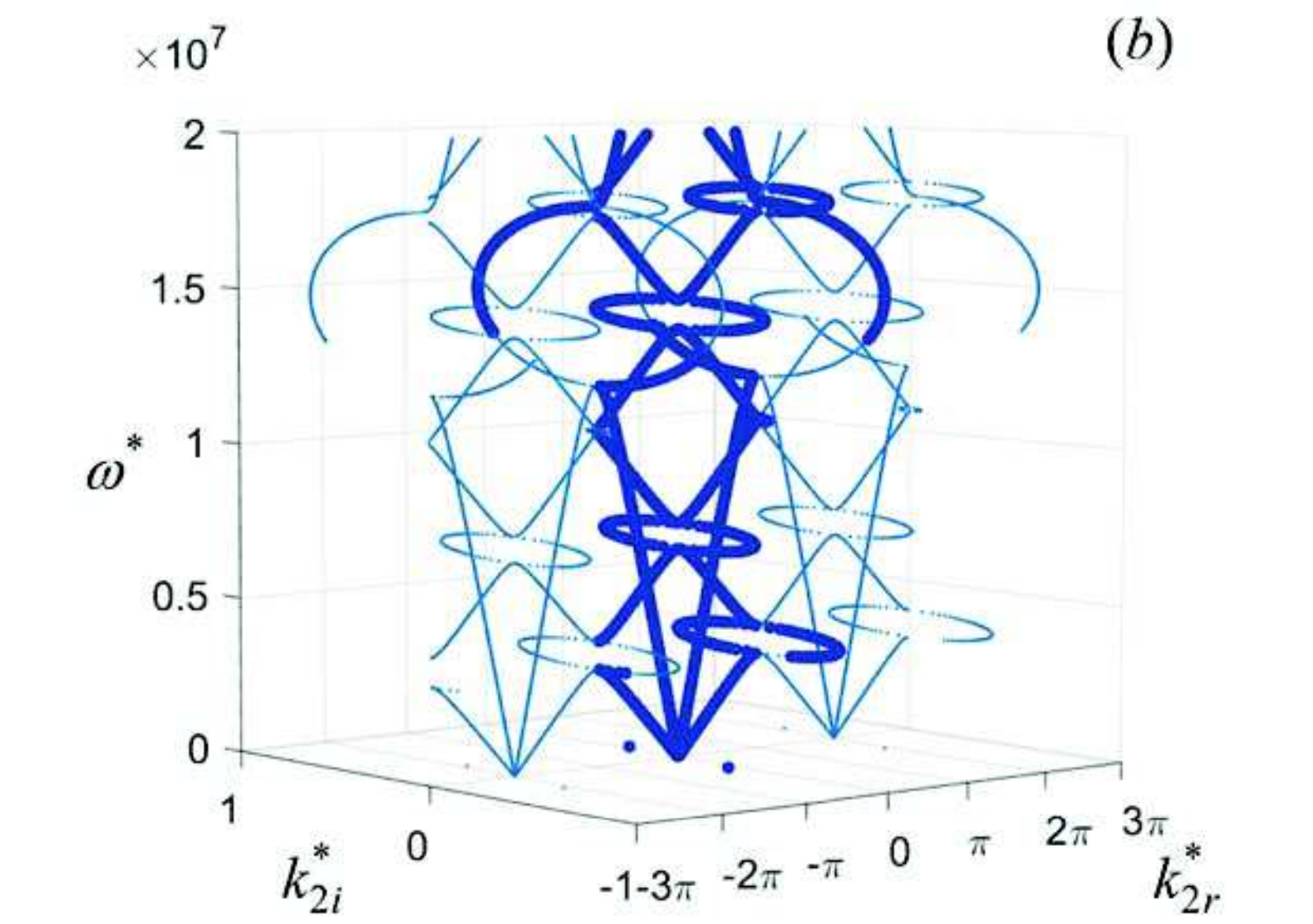}
  \\
 \hspace{-0.75cm}
  \includegraphics[width=6cm,angle=-90]{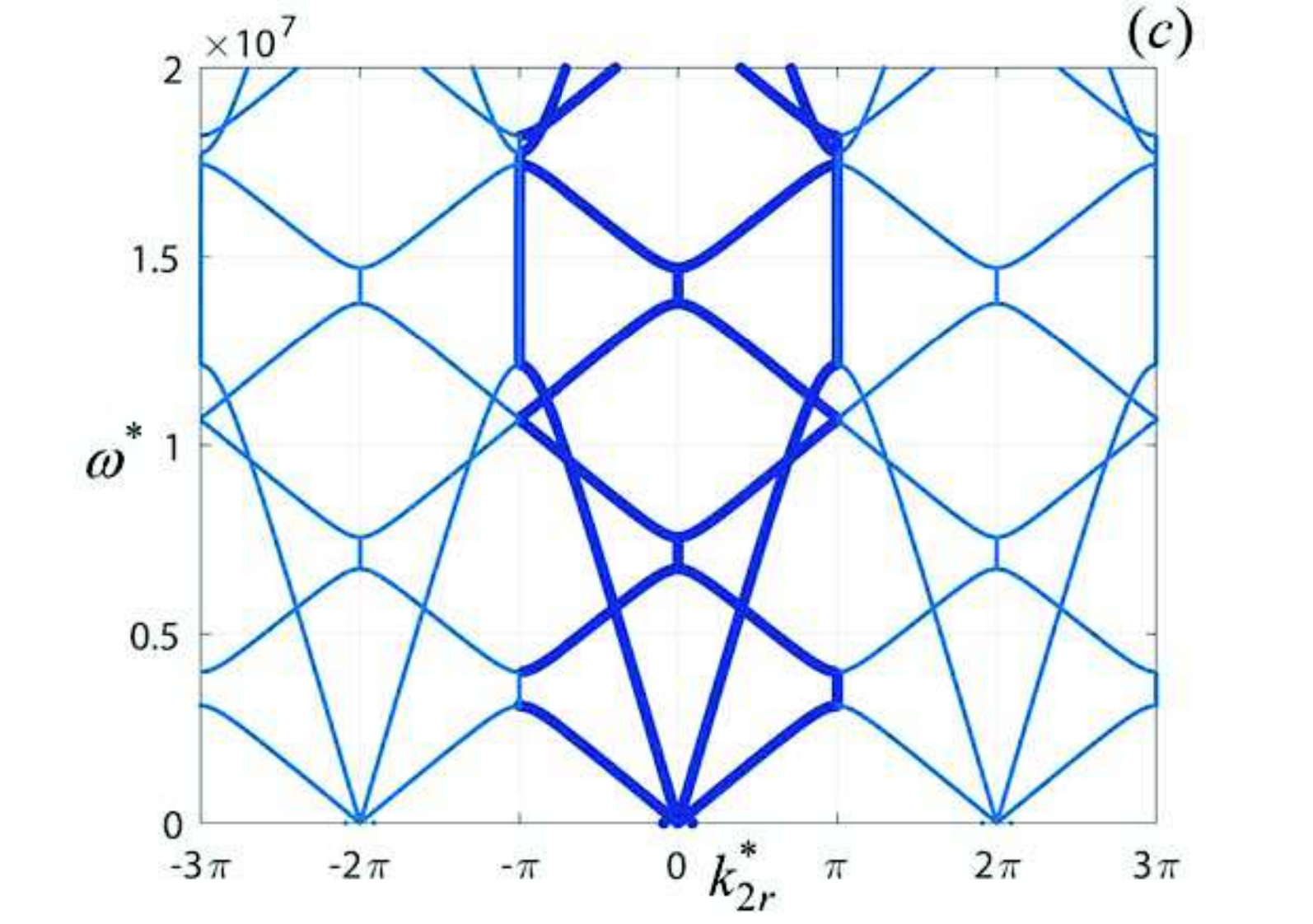}
  &
 \hspace{-1cm}
  \includegraphics[width=6cm,angle=-90]{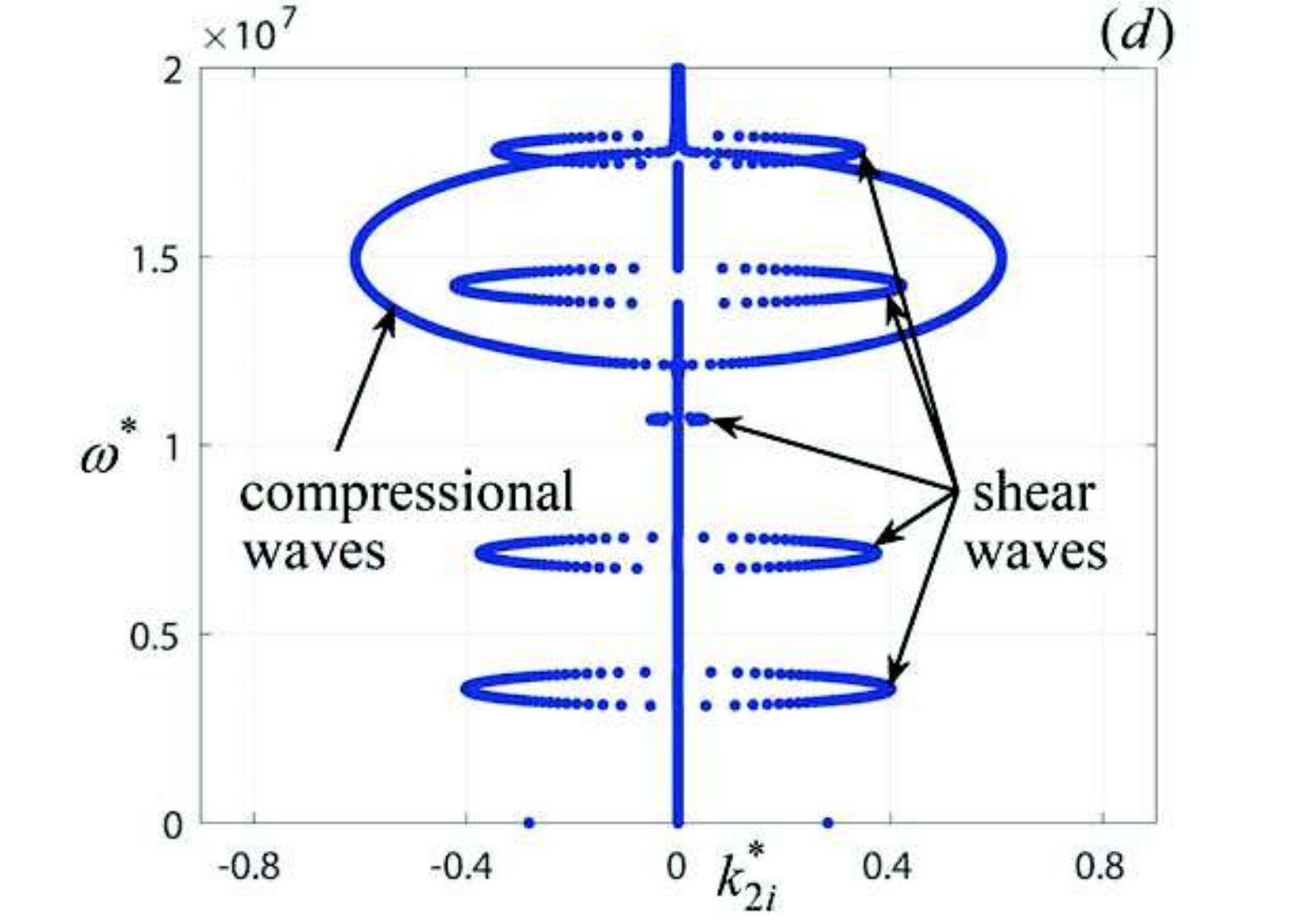}
  \\
 \hspace{-0.75cm}
  \includegraphics[width=8cm]{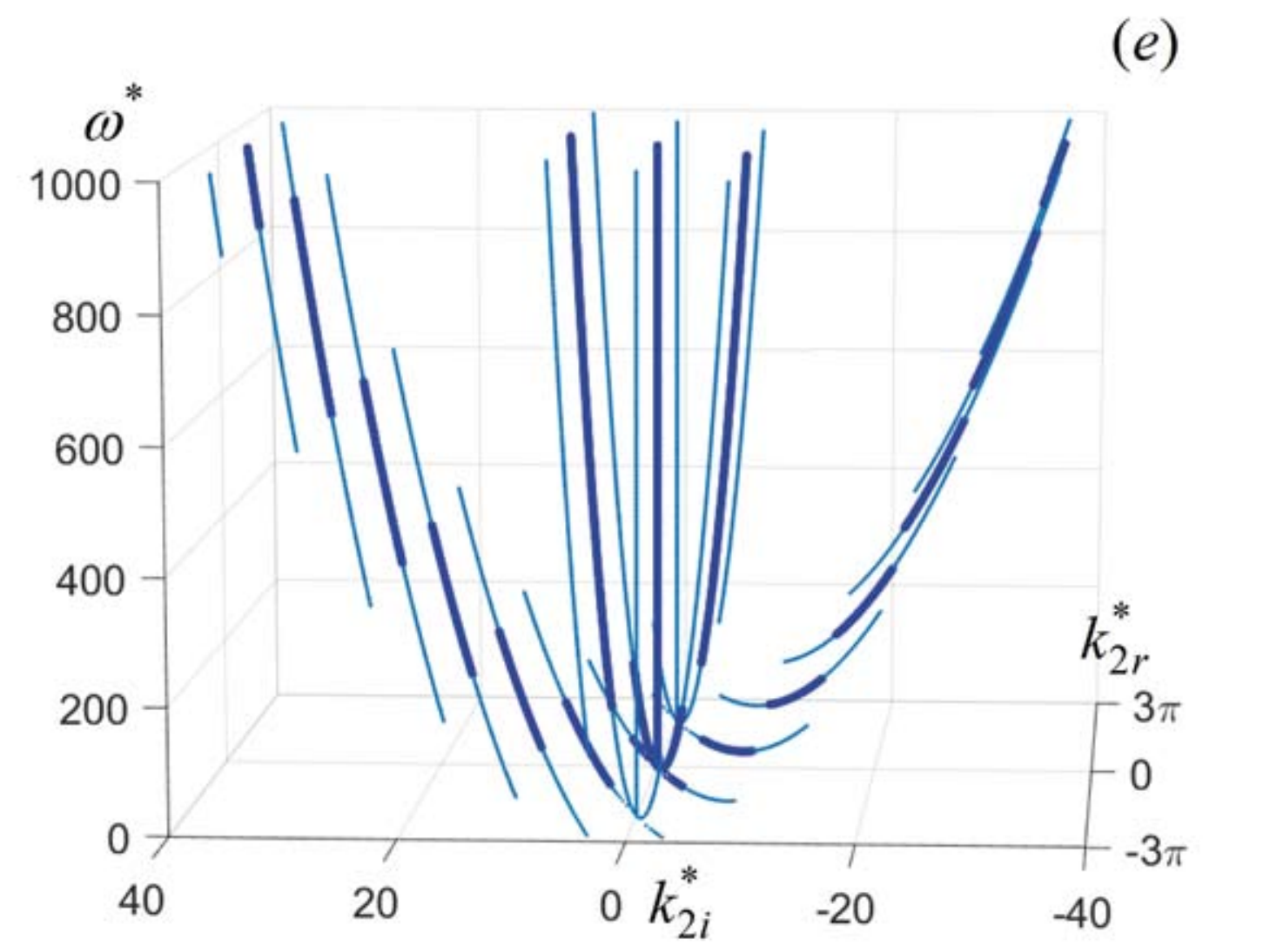}
  &
 \hspace{-1cm}
  \includegraphics[width=8cm]{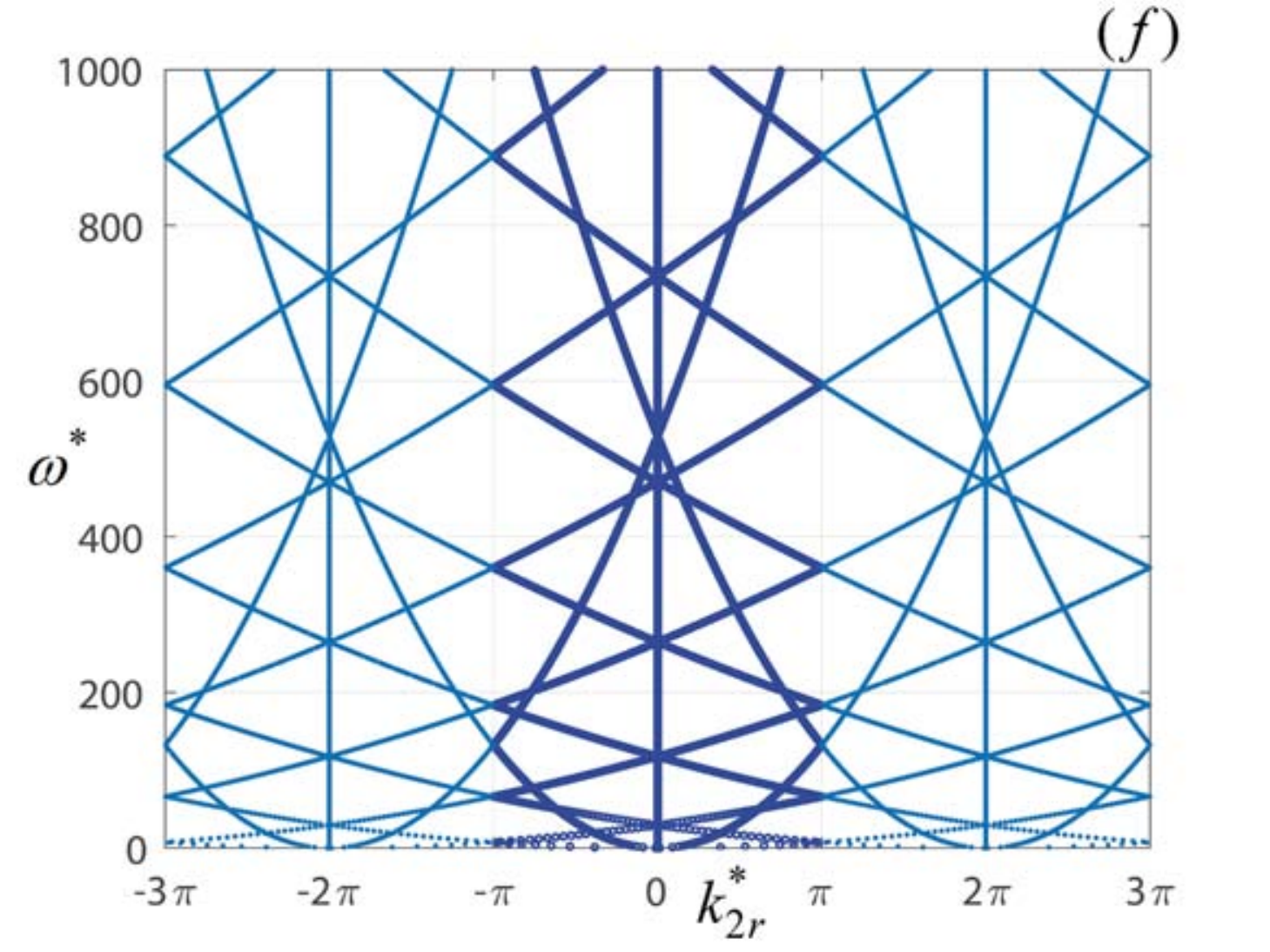}
  \end{tabular}
  \caption{Complex frequency spectrum obtained for $k_1=0$. (a)  3D view; (b) zoomed 3D view for $-1\leq k_{2i}^*\leq 1$; (c) plane $k_{2r}^*-\omega^*$ for $-1\leq k_{2i}^*\leq 1$; (d) plane $k_{2i}^*-\omega^*$ for $-1\leq k_{2i}^*\leq 1$; (e) 3D view for $0\leq \omega^*\leq 10^3$ and for $-40\leq k_{2i}^*\leq 40$; (f) plane $k_{2r}^*-\omega^*$ for $0\leq\omega^*\leq 10^3$. }
  \label{Fig::spectra1}
\end{figure}
For each phase, matrices $\mathbf{A},\mathbf{B}$, and $\mathbf{C}$ introduced in equation (\ref{ODEs}) assume the form
\begin{eqnarray}
\bA&=&
\left(
\begin{array}{cccc}
 G & 0 & 0 & 0 \\
 0 &  \cfrac{2G(1-\nu)}{1-2\nu} & 0 & 0 \\
 0 & 0 & K & 0 \\
 0 & 0 & 0 & D
\end{array}
\right), 
\nonumber\\
\bB&=& 
\left(
\begin{array}{cccc}
 2ik_{2}G & \cfrac{ik_{1}G}{1-2\nu} & 0 & 0 \\
  \cfrac{ik_{1}G}{1-2\nu} &  \cfrac{4ik_{2}G(1-\nu)}{1-2\nu} & -\alpha & -\beta \\
 0 & i\omega\alpha & 2ik_{2}K & 0 \\
 0 & i\omega\beta & 0 & 2ik_{2}D
\end{array}
\right),
\nonumber\\
 \bC&=& 
\left(
\begin{array}{cccc}
\left(\begin{array}{c}
\rho\omega^2\\
-\cfrac{2k_1^2G(1-\nu)}{1-2\nu}\\
-Gk_2^2
\end{array}\right)  & -\cfrac{k_1 k_2 G}{1-2\nu} & -ik_1\alpha & -ik_1\beta \\
  -\cfrac{k_1 k_2 G}{1-2\nu} &\left(\begin{array}{c}
  \rho\omega^2\\
   -k_1^2G\\
   -\cfrac{2k_2^2G(1-\nu)}{1-2\nu}
\end{array}\right)    & -ik_2\alpha & -ik_2\beta \\
 -\omega k_1 \alpha & -\omega k_2 \alpha &\left(\begin{array}{c}
 i\omega p\\
 -(k_1^2+k_2^2) K
\end{array}\right)   & i\omega\psi \\
 -\omega k_1 \beta & -\omega k_2 \beta & i\omega\psi & \left(\begin{array}{c}
 i\omega q\\
 -(k_1^2+k_2^2) D
 \end{array}\right)
\end{array}
\right). 
\nonumber\\
\label{mCex}
\end{eqnarray}
Figure \ref{Fig::spectra1} represents the complex frequency spectrum obtained by solving standard eigenvalue problem (\ref{eq:FBsyst}) in the direction perpendicular to the material layering ($k_1=0$).
In this case, the plane wave propagating inside the material results to be homogeneous since $\mathbf{n}_r\equiv\mathbf{n}_i$ in equation (\ref{eq:wavevector}).
Complex-valued wave number $k_2$ has been determined for discrete values of the real-valued frequency $\omega$ in a selected range, spanning from 0 to $2\cdot10^7$ rad/s.
Figure \ref{Fig::spectra1}-(a)
 plots the real and imaginary parts of wave number $k_2$, related to the complex-valued eigenvalue $\lambda$ through equations (\ref{eq:wavenumberk2}), in terms of $\omega$.
 In particular, real and imaginary parts of dimensionless wave number $k_2^*=k_2\,L$ are plotted in terms of  the real dimensionless frequency $\omega^*=\omega/\omega_{ref}$, being $\omega_{ref}=1\,\textrm{rad}/s$ a reference frequency.
 MATLAB$^{\circledR}$ enhanced with the Advanpix Multiprecision Toolbox has been exploited as a tool  for computing transfer matrix $\mathbf{T}_{(1,n)}$ of the periodic cell and solving linear eigenvalue problem (\ref{eq:FBsyst}).
 {The above mentioned toolbox allows computing using an arbitrary precision that, with respect to the usual double one, revealed to be an essential feature in order to obtain a unitary determinant for the symplectic matrix $\mathbf{T}_{(1,n)}$ and to compute the right eigenvalues.
 Involved matrices, in fact, are characterized by entries having absolute values that differ by several orders of magnitude. 
 The main practical difficulty  in finding the eigenvalues is that the eigenproblem might result  ill-conditioned and hard to compute. In this regard, using an arbitrary precision has been crucial in order to solve problem (\ref{eq:FBsyst}).} 
 Light blue curves of figure \ref{Fig::spectra1} represent the translation of the spectrum along the $k_{2r}^*$ axis in order to emphasize the periodicity of the curves along this axis.
 Figure \ref{Fig::spectra1}-(b) is a zoom of figure \ref{Fig::spectra1}-(a) considering $-1\leq k_{2i}^*\leq 1$, thus showing propagation branches related to the presence of hyperbolic equation (\ref{eq:FieldEquStress}) in the governing field equations set.
 Figures \ref{Fig::spectra1}-(c) and \ref{Fig::spectra1}-(d) are the two-dimensional representation of \ref{Fig::spectra1}-(b) displaying, respectively, the planes $k_{2r}^*-\omega^*$ and $k_{2i}^*-\omega^*$.

 \begin{figure}
  \centering
  \begin{tabular}{c c }
 \hspace{-0.75cm}
  \includegraphics[width=8cm]{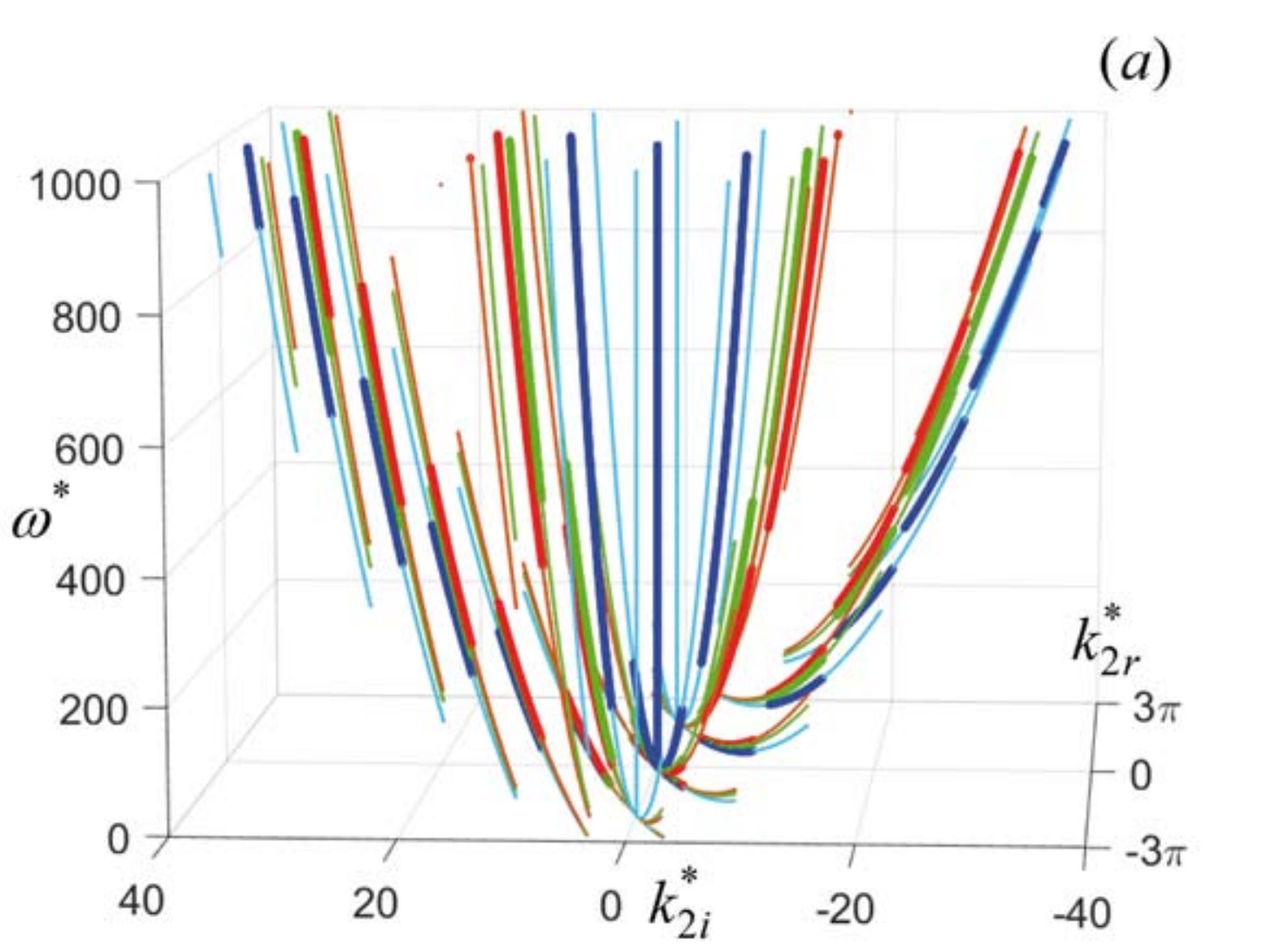}
  &
 \hspace{-1cm}
  \includegraphics[width=8cm]{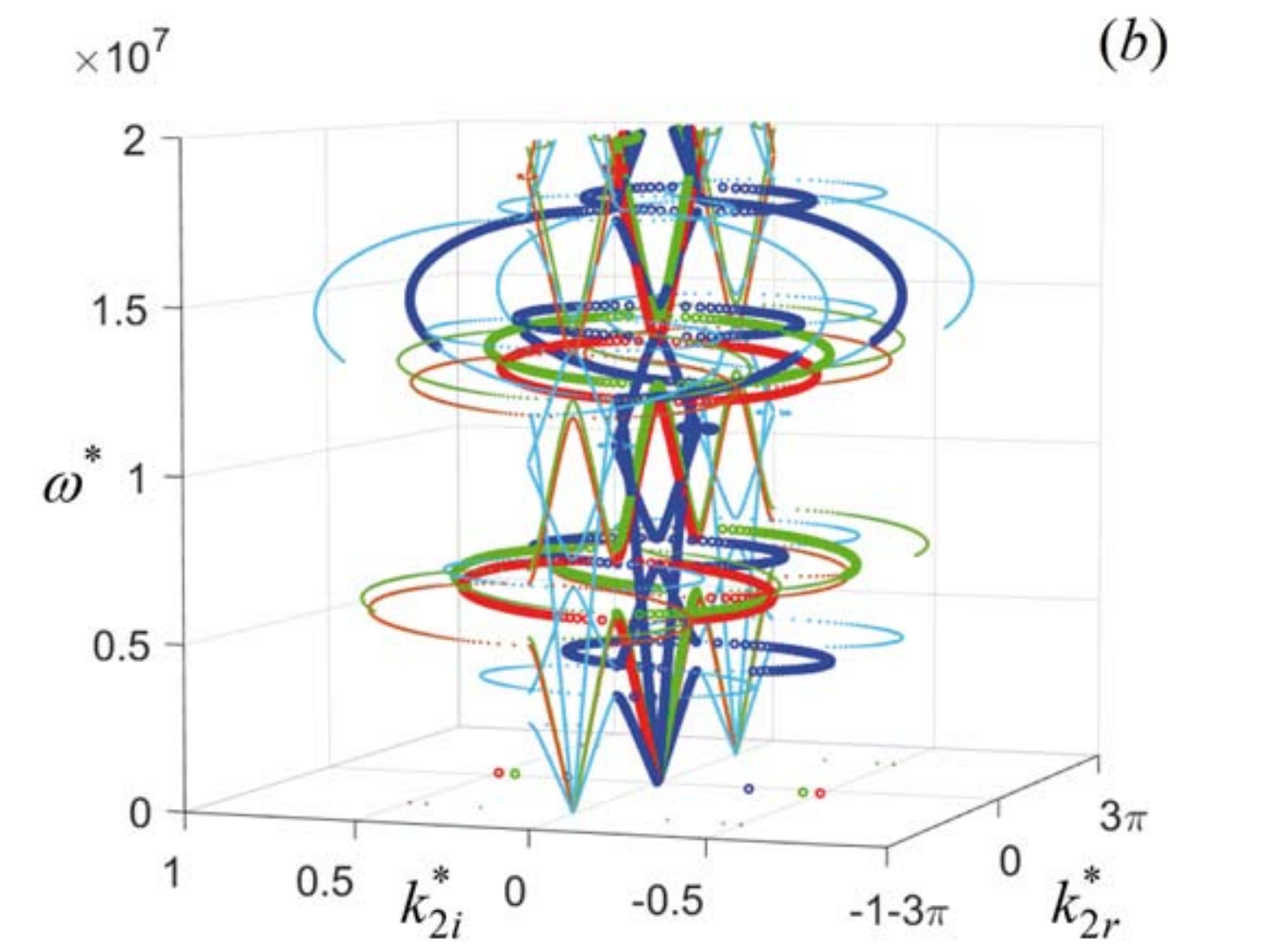}
  \\
 \hspace{-0.75cm}
  \includegraphics[width=8cm]{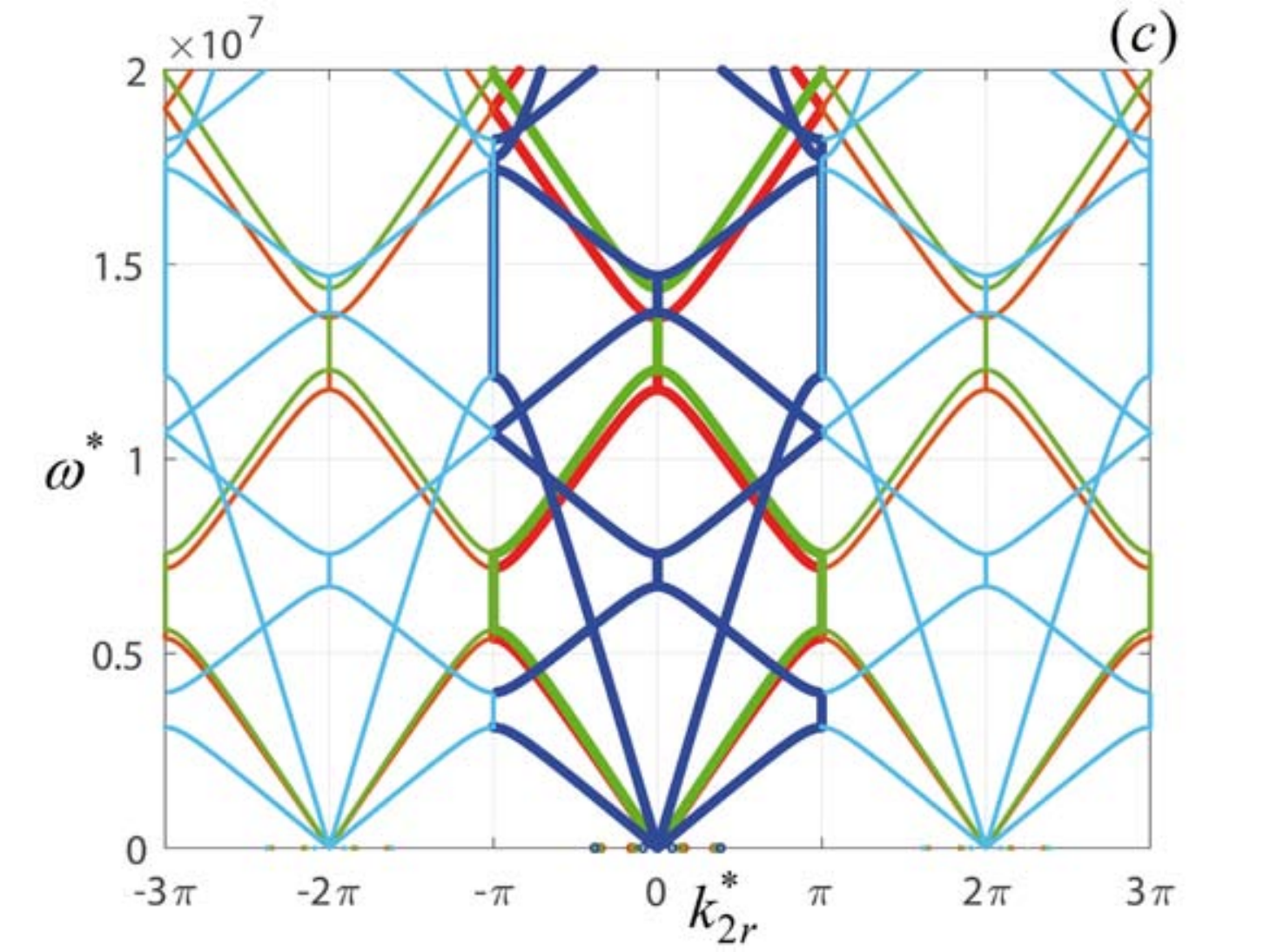}
  &
 \hspace{-1cm}
 \includegraphics[width=8cm]{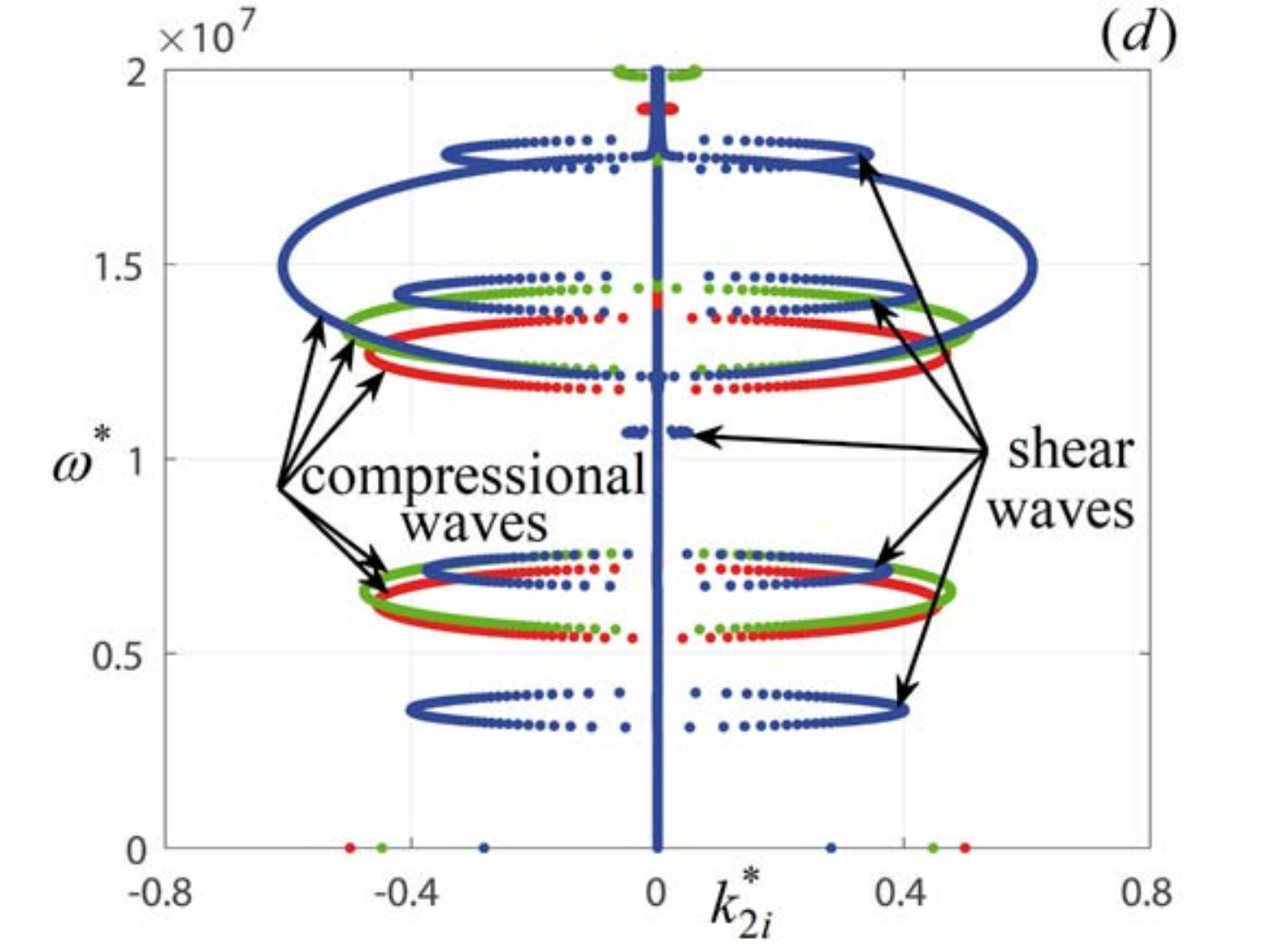}
  \\
 \hspace{-0.75cm}
  \includegraphics[width=8cm]{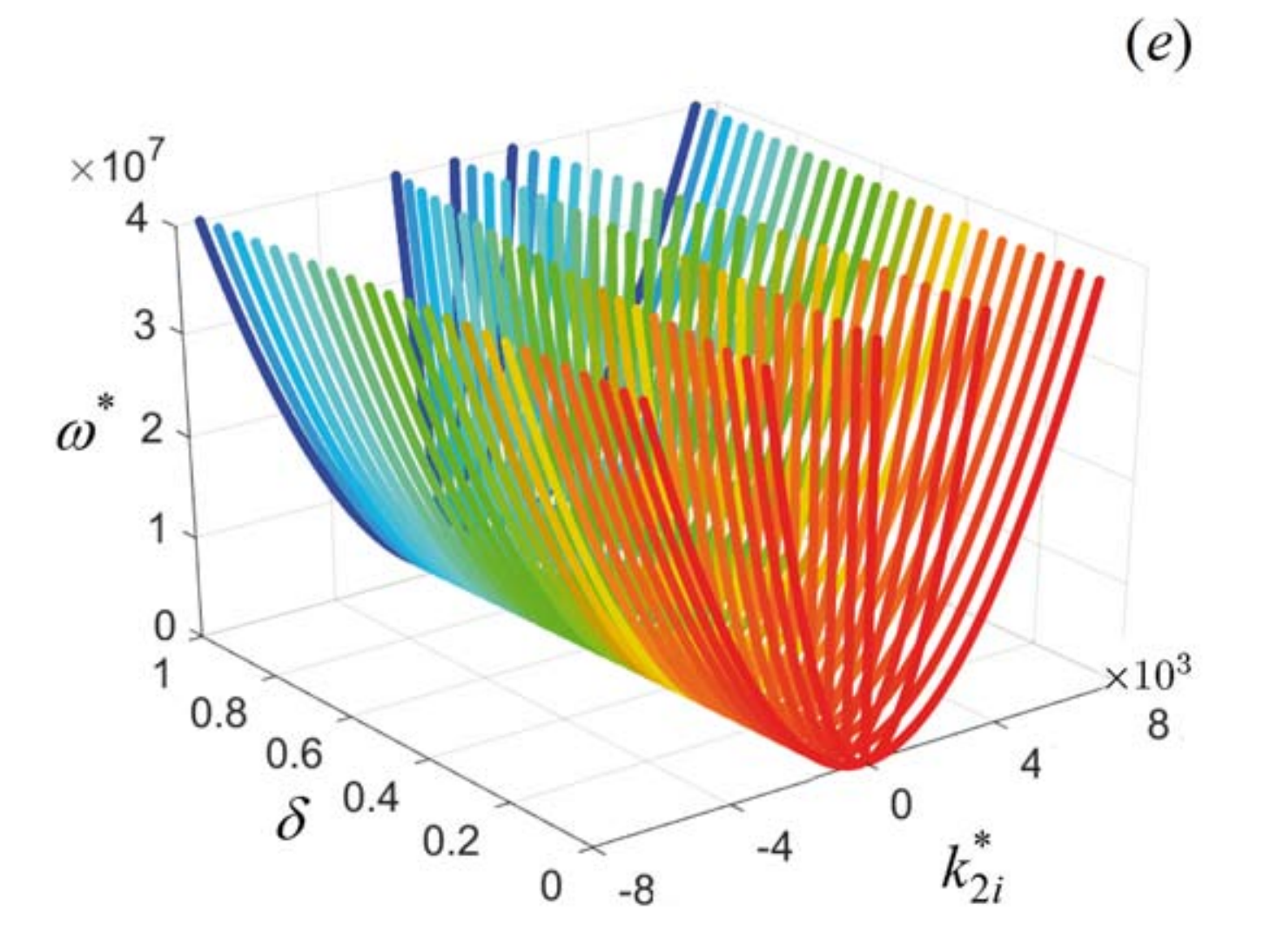}
  &
 \hspace{-1cm}
  \includegraphics[width=8cm]{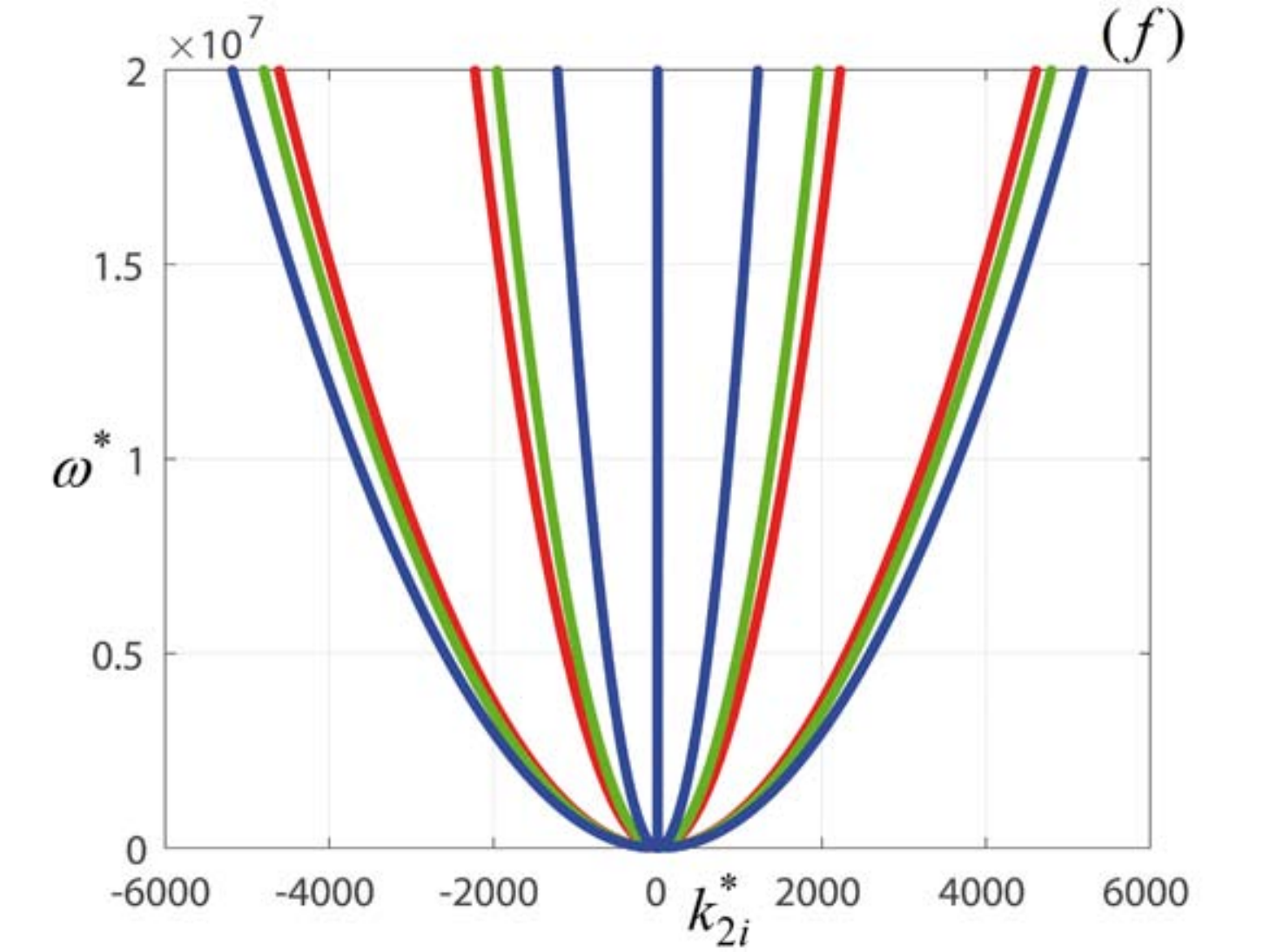}
  \end{tabular}
  \caption{Complex frequency spectrum obtained for $k_1=0$ and different values of the coupling factor: $\delta=0$ (red curves), $\delta=0.5$ (green curves), and $\delta=1$ (blue curves). (a) 3D view zoomed for $0\leq \omega^*\leq 10^3$; (b) 3D view zoomed for $-1\leq k_{2i}^*\leq 1$; (c) plane $k_{2r}^*-\omega^*$  for $-1\leq k_{2i}^*\leq 1$; (d) plane $k_{2i}^*-\omega^*$  for $-1\leq k_{2i}^*\leq 1$; (e) $k_{2i}^*$ as a function of $\delta$ and $\omega^*$; (f) plane $k_{2i}^*-\omega^*$. }
  \label{Fig::spectra2}
\end{figure}
\begin{figure}
  \centering
  \begin{tabular}{c c }
 \hspace{-0.75cm}
  \includegraphics[width=8cm]{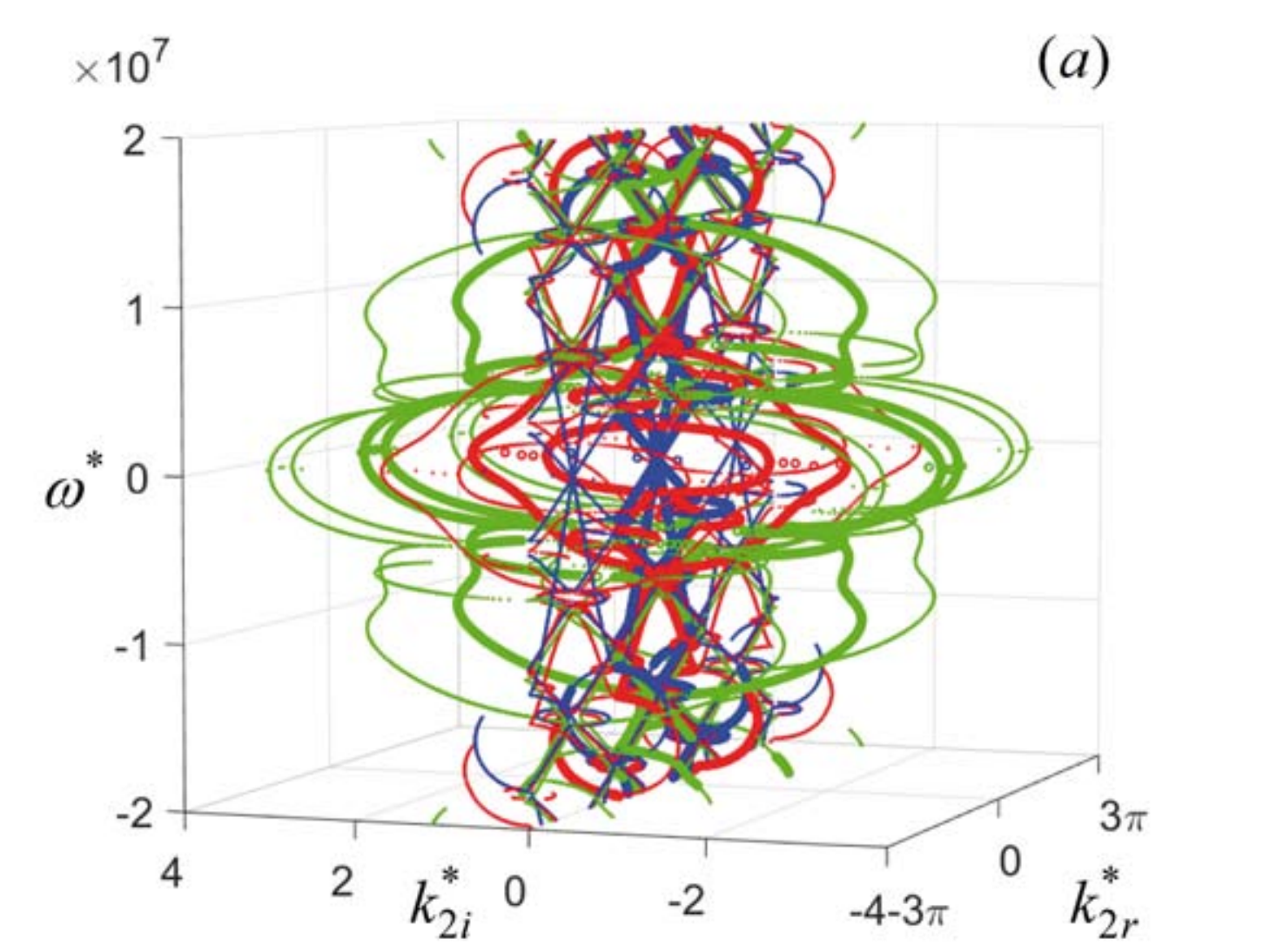}
  &
 \hspace{-1cm}
  \includegraphics[width=8cm]{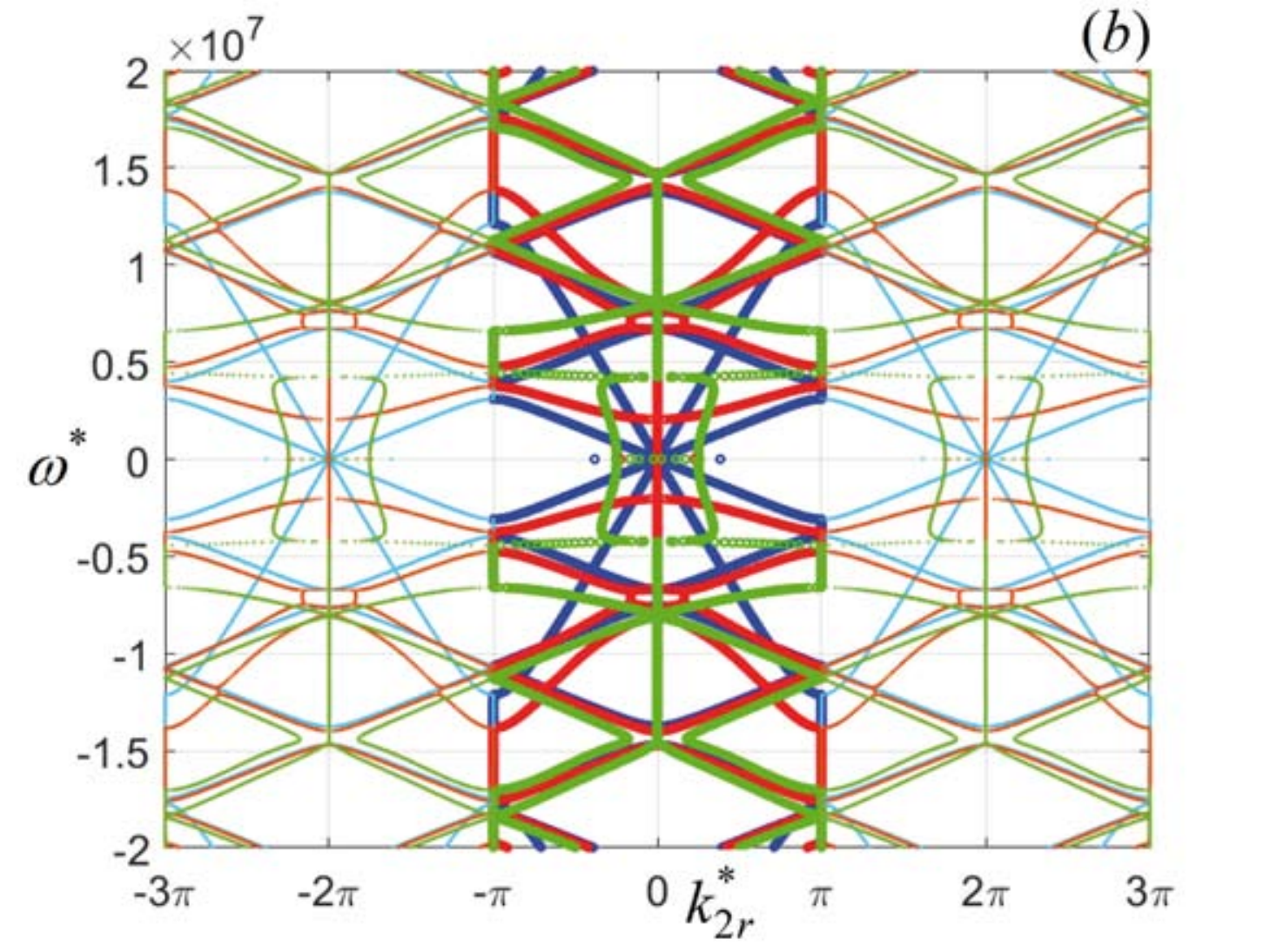}
  \\
 \hspace{-0.75cm}
  \includegraphics[width=8cm]{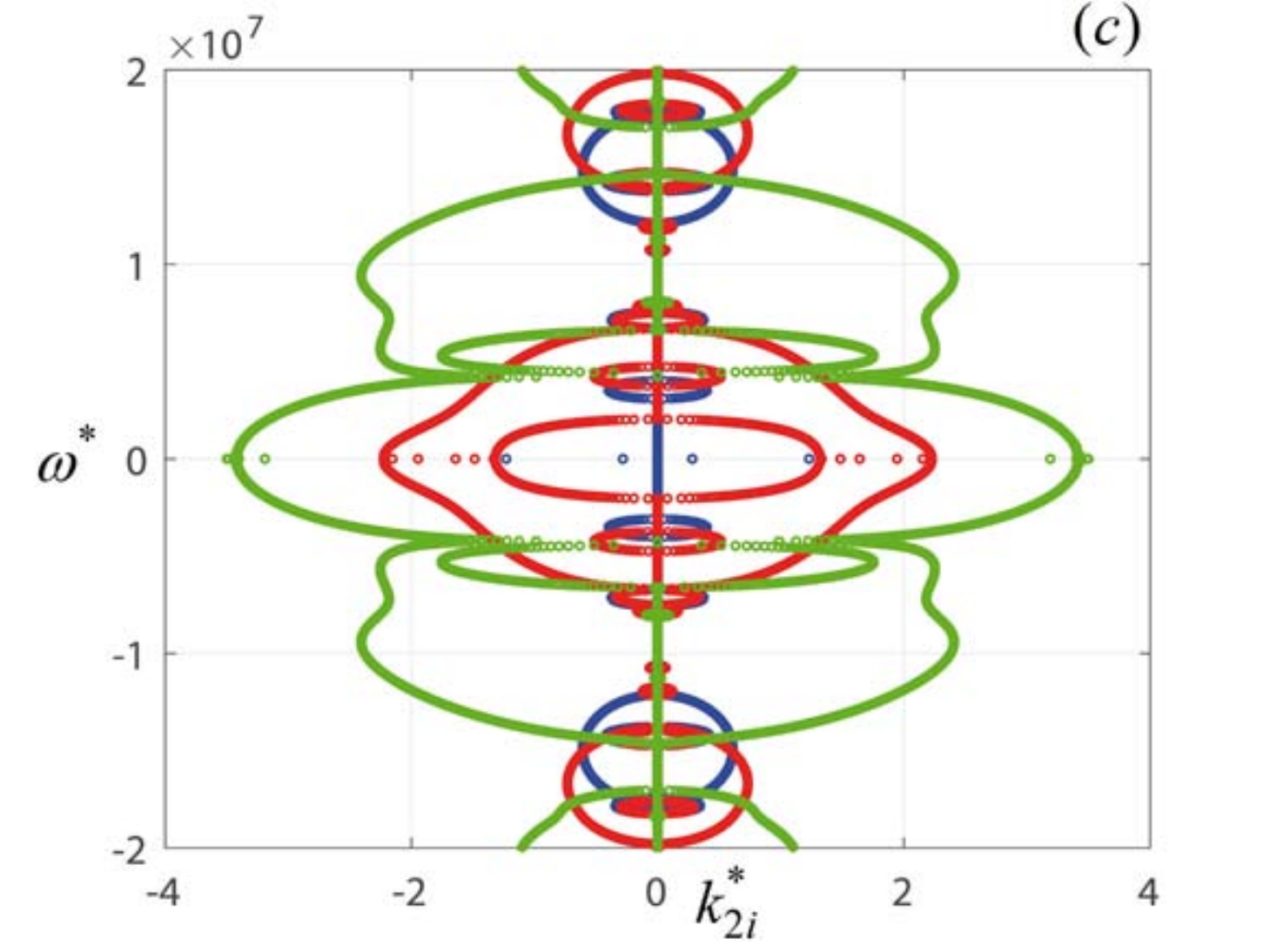}
  &
 \hspace{-1cm}
 \includegraphics[width=8cm]{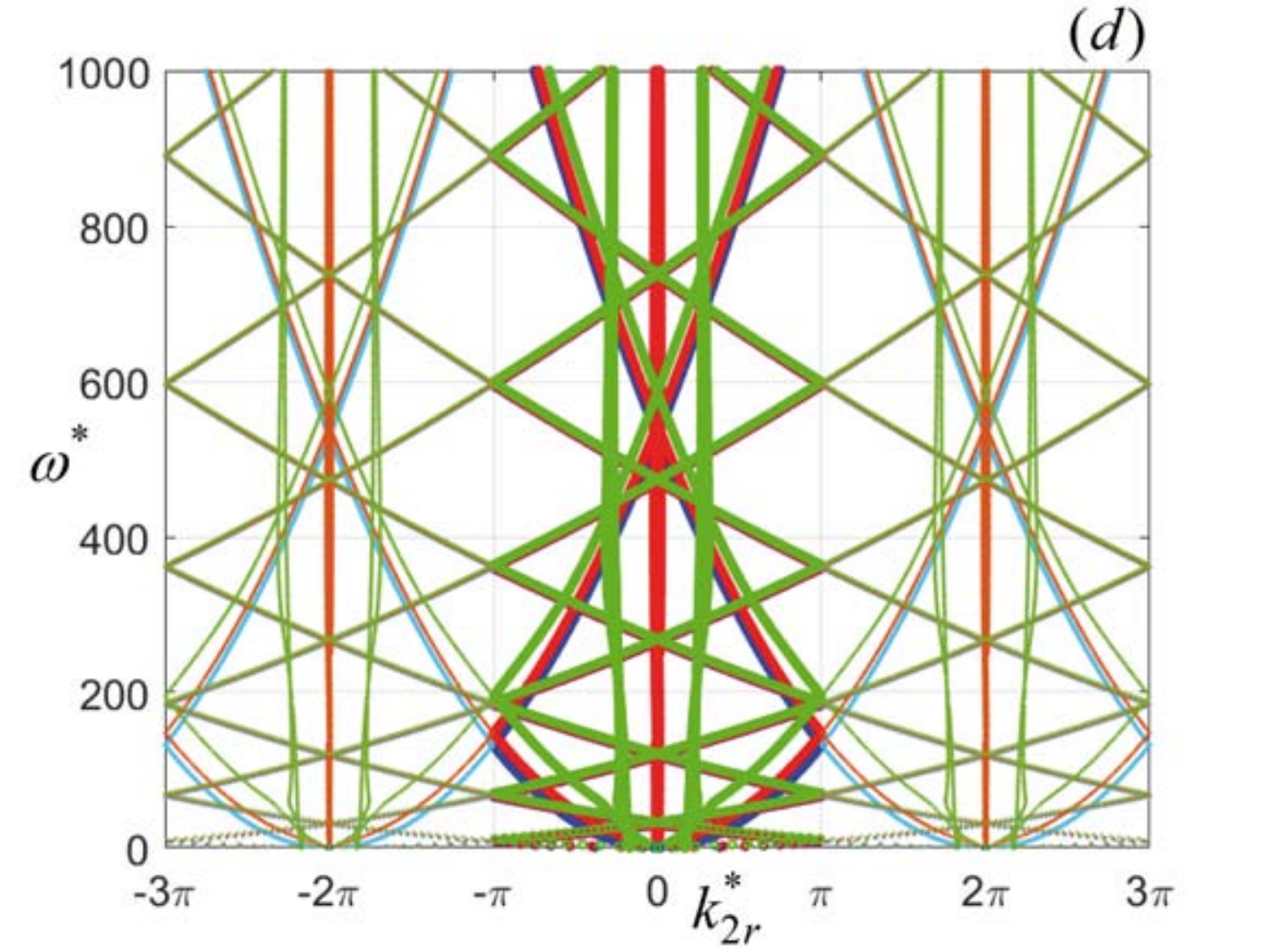}
  \end{tabular}
  \caption{ Complex material spectra obtained for $k^*_1=0$ (blue curves), $k_1^*=0.5\,\pi$ (red curves), $k^*_1=\pi$ (green curves). (a) 3D view for $-4\leq k_{2i}^* \leq 4$; (b) plane $ k_{2r}^*-\omega^*$ for $-4\leq  k_{2i}^* \leq 4$; (c) plane $ k_{2i}^*-\omega^*$ for $-4\leq  k_{2i}^* \leq 4$; (d) plane $ k_{2r}^*-\omega^*$ for $0\leq \omega^*\leq 10^3$. }
  \label{Fig::spectra3}
\end{figure}

 They show, respectively, the structure of pass bands with real-valued wave number $k_2^*$ corresponding to propagating waves, and the structure of band gaps with
 imaginary wave number $k_2^*$, which describes spatial wave attenuation due to material damping.
 Figure \ref{Fig::spectra1}-(d) clearly plots the opening of different band gaps, related to both  compressional and shear mechanical waves, where the second ones result to be uncoupled  from thermal and diffusive fields being components  $\alpha_{12}$ and $\beta_{12}$ of constitutive tensors $\bga$ and $\bgb$, respectively, equal to zero for both phases of the unit cell.
 Figure \ref{Fig::spectra1}-(e) is a zoomed view of figure \ref{Fig::spectra1}-(a) with $0\leq\omega^*\leq 10^3$ detailing the behaviour of damping branches due to the existence of the two parabolic equations (\ref{eq:FieldEquThermalFlux}) and (\ref{eq:FieldEquMassFlux}) in the governing field equations set, which give rise to the two parabolas in the plane $k_{2i}^*-\omega^*$.
 Figure \ref{Fig::spectra1}-(f) is the two-dimensional representation of figure \ref{Fig::spectra1}-(e) in the plane $k_{2r}^*-\omega^*$. 
{It is here anticipated that the two-dimensional representation of figure \ref{Fig::spectra1}-(e) in the plane $k_{2i}^*-\omega^*$ corresponds to the blue curves represented in figure \ref{Fig::spectra2}-(f).}
Figure \ref{Fig::spectra2} shows the changes that occur in the material band diagrams because of variations in the values of thermodiffusive coupling, again  in the case $k_1=0$.
In particular, premultiplying $\bga$, $\bgb$ and $\psi$ in equations (\ref{eq:FieldEquStress})-(\ref{eq:FieldEquMassFlux}) by a scalar coupling factor $\delta$, blue curves of figure \ref{Fig::spectra2}
 represent the case $\delta=1$, green curves the case $\delta=0.5$, and red curves the case $\delta=0$, this last corresponding to the fully uncoupled state.
 As in figure \ref{Fig::spectra1}, obtained spectra have been translated along the $k^*_{2r}$ axis using, for each value of $\delta$, a thin and  light marker, in order to stress the periodicity of the curves along that axis.
Figure \ref{Fig::spectra2}-(a) is a three-dimensional representation of computed band diagrams for $0\leq\omega^*\leq 10^3$ showing the behaviour of damping branches.
Figure \ref{Fig::spectra2}-(b) is a zoomed view of the three-dimensional spectra for $-1\leq k^*_{2i}\leq 1$ depicting the behaviour of propagation branches and figures \ref{Fig::spectra2}-(c) and \ref{Fig::spectra2}-(d) are its corresponding two-dimensional representations, respectively in the plane $k^*_{2r}-\omega^*$ and $k^*_{2i}-\omega^*$. 
As expected, pass bands and band gaps structure of shear waves is not influenced by the value of the coupling factor $\delta$, being mechanical shear waves uncoupled from thermal and diffusive fields, while the behaviour of compressional waves results strongly affected by thermodiffusive coupling.
In particular, figure \ref{Fig::spectra2}-(c) shows a broadening of pass bands width as $\delta$ increases, with a consequent increase of the mean frequency value of each pass band.
On the other hand, figure \ref{Fig::spectra2}-(d) exhibits a broadening of band gaps width as the coupling factor increases, which is a desirable feature for different frequency sensing and noise isolation applications.
Furthermore, the mean frequency value of each band gap increases as $\delta$  increases.
Figure \ref{Fig::spectra2}-(e) is a three-dimensional representation of the imaginary part of the wave number $k^*_{2i}$ in terms of $\delta$ and $\omega^*$, showing the influence of thermodiffusive coupling upon the behaviour of  damping branches.
As clearly represented also in figure \ref{Fig::spectra2}-(f), which is a two-dimensional representation of figure \ref{Fig::spectra2}-(e) in the plane $k^*_{2i}-\omega^*$ for three selected values of the coupling factor ($\delta=0$, $\delta=0.5$, and $\delta=1$), the external parabolas increase their amplitudes as  $\delta$ increases, which corresponds, for the same value of frequency $\omega^*$, to a higher spatial attenuation ($k^*_{2i}$ positive) or amplification ($k^*_{2i}$ negative) of the wave as thermodiffusive coupling increases.
On the contrary, internal parabolas decrease their amplitudes as $\delta$ increases, with a consequent decreasing of the spatial attenuation/amplification of the wave as $\delta$ increases for each value of the frequency $\omega^*$.
\begin{figure}[h!]
  \centering
  \begin{tabular}{c c }
 \hspace{-0.75cm}
  \includegraphics[width=6cm,angle=-90]{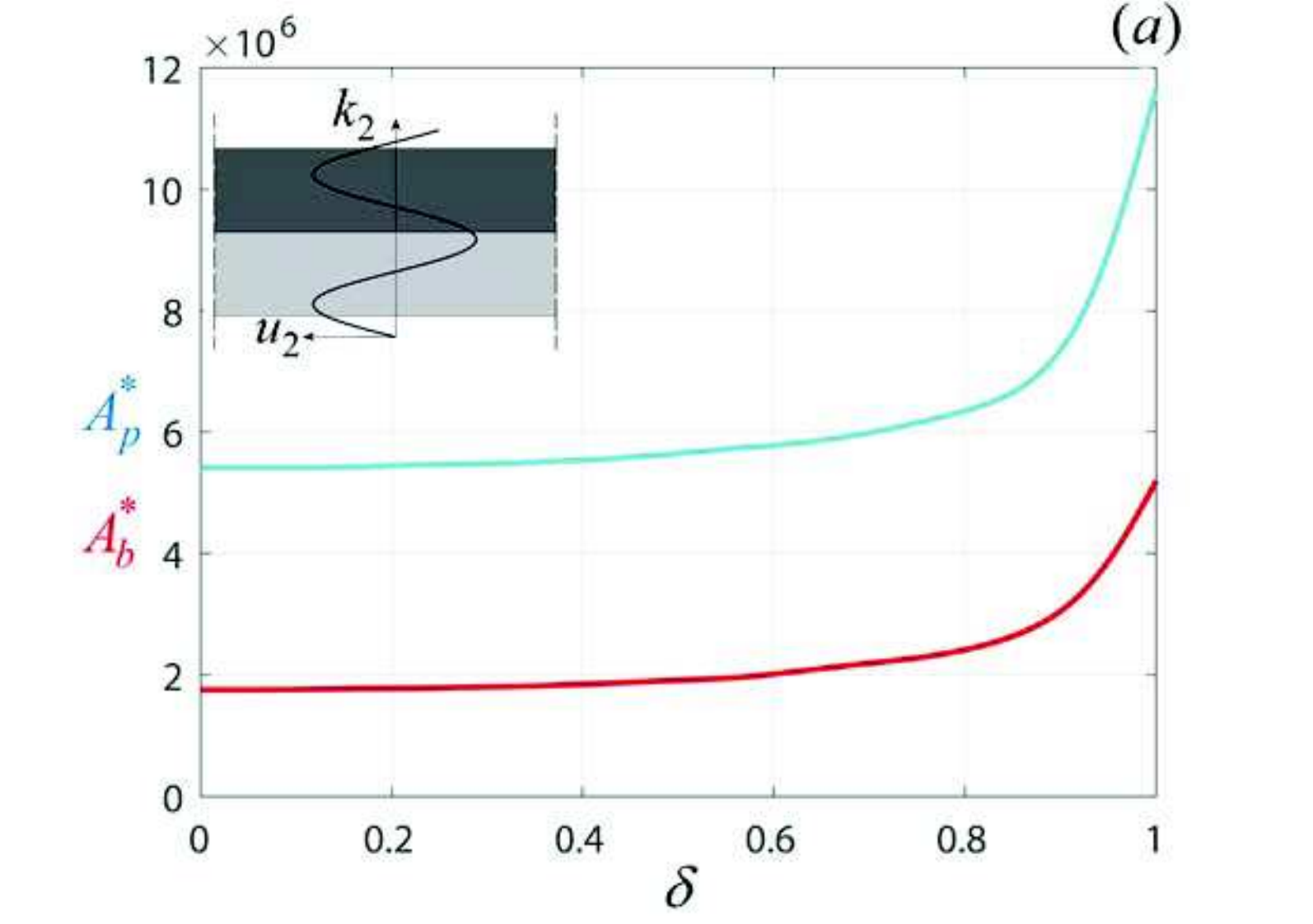}
  &
 \hspace{-1cm}
  \includegraphics[width=6cm,angle=-90]{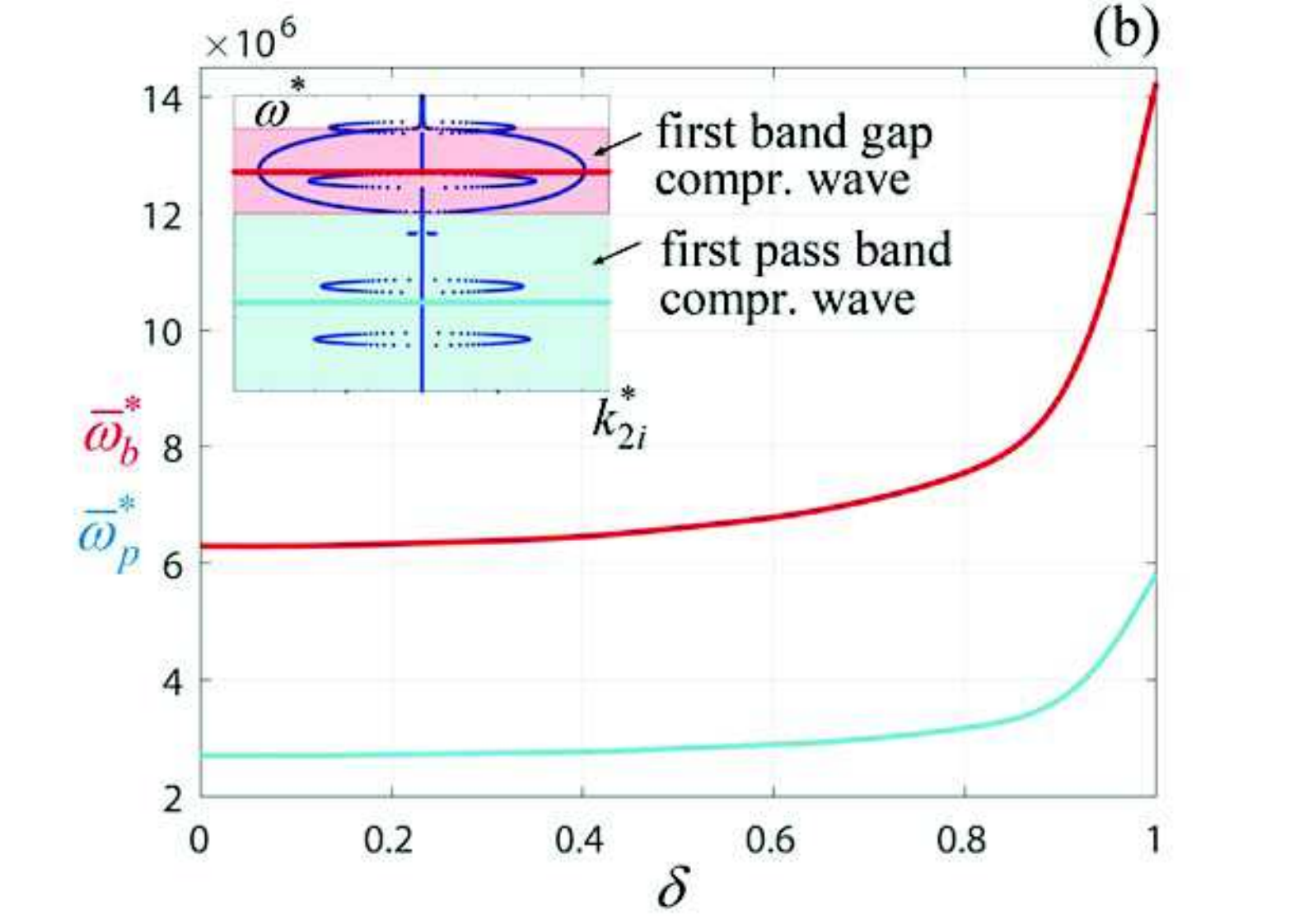}
  \end{tabular}
  \caption{(a)  Dimensionless width of the first pass band $A^*_p$ (light blue curve) and of the first band gap $A^*_b$ (red curve) relative to compressional waves vs coupling factor $\delta$;   (b) Dimensionless mean frequency of the first pass band $\bar{\omega}^*_p$ (light blue curve) and of the first band gap $\bar{\omega}^*_b$ (red curve) relative to compressional waves vs coupling factor $\delta$. }
  \label{Fig::spectra2b}
\end{figure}
Figure \ref{Fig::spectra2b} stresses the influence of thermodiffusive coupling upon the behaviour of the first pass band and of the first band gap for compressional waves.
In particular, figure \ref{Fig::spectra2b}-(a) depicts the increase of the width of the first pass band $A^*_p$ (light blue curve) and of the fist band gap $A^*_b$ (red curve) as $\delta$ increases, while figure \ref{Fig::spectra2b}-(b) shows the increase of the mean frequency value relative to the first pass  band $\bar{\omega}^*_p$ (light blue  curve) and to the first band gap $\bar{\omega}^*_b$ (red  curve) in terms of the coupling factor $\delta$.
Both widths and  mean frequencies have been adimensionalized with the reference frequency $\omega_{ref}$.
Finally, figure \ref{Fig::spectra3}
 refers to spectra obtained  for different values of dimensionless wave number $k_1^*=k_1\,L$, assumed to have a vanishing imaginary component.
 Blue curves denote the case $k_1^*=0$, red curves the case $k_1^*=0.5\pi$, and green curves the case $k_1^*=\pi$.
 Figure \ref{Fig::spectra3}-(a) is a section in $\mathbb{R}^3$ of the hypercurves described in  (\ref{eq:Hypersurfaces}) for $-4\leq k_{2i}^*\leq 4$, showing propagation branches related to the hyperbolic equations (\ref{eq:FieldEquStress}) in the governing field equations set.
 %
 %
 %
 %
 Figures \ref{Fig::spectra3}-(b) and \ref{Fig::spectra3}-(c) show, respectively, the two-dimensional representations of figure \ref{Fig::spectra3}-(a) in the planes $k_{2r}^*-\omega^*$ and $k_{2i}^*-\omega^*$.
 Figure \ref{Fig::spectra3}-(d) is a zoomed view of obtained spectra in the plane $k_{2r}^*-\omega^*$ for $0\leq \omega^*\leq 10^3$, illustrating the behaviour of damping branches related to the presence of parabolic equations (\ref{eq:FieldEquThermalFlux})-(\ref{eq:FieldEquMassFlux}) in the governing field equations set. 
It is worth noting that plots in figure \ref{Fig::spectra3} are not sufficient in order to investigate the behaviour of a wave propagating inside the thermodiffusive composite material along directions different from the one that is perpendicular to material layering, for which both $k_{2}$ and $k_1$ vary  point by point.
They represent obtained complex spectra for a fixed value of wave number $k_1$, that, when is different from zero, characterizes the plane wave as inhomogeneous, since $\mathbf{n}_r\neq \mathbf{n}_i$ in equation (\ref{eq:wavevector}).

\section{Conclusions}
\label{eq:Conclusions}
The present work is devoted to investigate the propagation and damping of waves inside composite materials whose phases can be modeled as linear thermodiffusive media.
The principal goal is  the study and the estimation of the impact that thermal and diffusive effects can have upon the propagation of harmonic oscillation in two-dimensional thermodiffusive laminates. %
Materials frequency band structure and relative dispersion curves are provided in the case of complex-valued wave vectors and real angular frequencies (spatial damping), both for the uncoupled and coupled case and the changes  observed in the frequency spectra due to thermodiffusive couplings are discussed in details.
In the formulation, elastic wave equation is coupled with standard heat conduction and mass diffusion  equations, these lasts both of parabolic type and associated to damping phenomena.
In order to build material band diagrams,  after fixing the value of the wave number in the direction parallel to material layering, a standard eigenvalue problem is solved  in terms of the Floquet multiplier by spanning a selected range of frequency, here considered as an independent parameter.
Real and imaginary part of the wave number in the direction perpendicular to material layering, which are related, respectively, to the propagation and spatial attenuation (or amplification) of the wave, are then computed from the obtained  values of the complex Floquet multiplier.
Characteristic polynomial  valid for a  periodic thermodiffusive laminate, whose elementary cell is considered as made by an arbitrary number of layers, has been obtained by means of a generalization of the transfer matrix method  and by imposing generalized Floquet-Bloch  quasiperiodic conditions in the direction perpendicular to material layering. Floquet-Bloch approach, in fact, allows constructing a band diagram for an entire periodic medium by analyzing the dynamics of only a single unit cell.   
Illustrative examples are then provided, applying the developed general method to study the propagation and damping of harmonic   oscillations to bi-phase isotropic thermodiffusive laminates  of interest  for SOFCs applications. 
Vulnerability to damage of such devices can increase because  of typical high operating temperature and intensive ions flows and an accurate prediction of their performances reveals  to be of fundamental importance in order to not undermine their efficiency.
By varying the value of coupling terms in the governing field equations set, a broadening of band gaps widths associated to compressional waves  has been obtained as thermodiffusive coupling increases, which is a desirable feature in different isolation and sensing applications.
Furthermore, also the mean frequency value of pass bands and of band gaps  relative to mechanical compressional waves increases as the coupling increases.
Homogeneous and inhomogeneous waves have been investigated, depending  on whether the normals to planes having constant phase are parallel to normals to planes with constant amplitude or not.

\section*{Appendix A. Matrix exponential determination for a single layer of the composite laminate  }
General formal solution of  system \eq{ODEr} can be expressed in the form
\begin{equation}
\mathbf{r}=a \bgg\,\textrm{exp}\left[-\varsigma x_2\right],
\end{equation}
where $a$ is a constant, $\bgg$ is the eigenvector corresponding to the eigenvalue $\varsigma$, solution of the following associate eigenvalues problem
\begin{equation}
\mathbf{H}(\varsigma)\bgg=\mathbf{0},   
\label{eigensystem}
\end{equation}
with $\mathbf{H}(\varsigma)=\mathbf{N}-\varsigma\mathbf{M}$.
The existence of non-trivial solutions of the algebraic system \eq{eigensystem} requires the vanishing of the determinant of the matrix $\mathbf{H}$. This yields an eight-degree polynomial 
characteristic equation having the form
\begin{equation}
\mathcal{Q}(\varsigma)=\textrm{Det}\left(\mathbf{H}(\varsigma)\right)=\mathcal{Q}_{8}\varsigma^8+\mathcal{Q}_{6}\varsigma^{6}+\mathcal{Q}_{4}\varsigma^4+\mathcal{Q}_{2}\varsigma^2+\mathcal{Q}_{0}=0.
\label{Pgamma}
\end{equation}
The solution of equation \eq{Pgamma} gives the complete eigenvalues spectrum.
Assuming that this equation admits eight different solutions, and then that all eigenvalues are distinct, for each 
one of them one can determine the associate eigenvector $\bgg^{(i)}$ with $i=1,...,8$.
In this way, one obtains a complete set of eigenfunctions, which represents a basis of the solutions space,
and  the general solution can be written as a linear combination of these eigenfunctions
\begin{equation}
\mathbf{r}=\bgG\,\mathbf{E}\,\mathbf{a},
\label{gensol}
\end{equation}
where $\bgG=\left(\bgg^{(1)}\, \bgg^{(2)}\, \bgg^{(3)}\, \bgg^{(4)}\, \bgg^{(5)}\, \bgg^{(6)}\, \bgg^{(7)}\, \bgg^{(8)}\right)$ is the eigenvectors matrix with eigenvectors arranged by column, 
$\mathbf{a}=(a_1\,  a_2\,  a_3\,  a_4\,  a_5\,  a_6\,  a_7\,  a_8)^T$ is a constant vector, and $\mathbf{E}$ is a diagonal matrix of the form
\begin{eqnarray}
\mathbf{E}&=&\textrm{diag}\left[\textrm{exp}\left[-\varsigma^{(1)}x_2\right],\textrm{exp}\left[-\varsigma^{(2)}x_2\right],\textrm{exp}\left[-\varsigma^{(3)}x_2\right],\textrm{exp}\left[-\varsigma^{(4)}x_2\right],\textrm{exp}\left[-\varsigma^{(5)}x_2\right],\right.
\nonumber\\
&&\left.\textrm{exp}\left[-\varsigma^{(6)}x_2\right],\textrm{exp}\left[-\varsigma^{(7)}x_2\right],\textrm{exp}\left[-\varsigma^{(8)}x_2\right]\right].
\label{eq:matrixE}
\end{eqnarray}
Matrix $\mathbf{E}$ is diagonalizable when algebraic multiplicity of the eigenvalues equals their geometric multiplicity, otherwise $\mathbf{E}$ assumes the form of a Jordan block diagonal matrix. 
Note that assuming the form \eq{gensol} for the solution of  system \eq{ODEr} implies that all the eigenvalues $\gamma_{j}$ are distinct. 
If some eigenvalues are identical, the exponential matrix assumes a more complicated
form including terms depending by $x_2^{n}$,  where $n$ is the degree of degeneracy of the system \citep{Arfk1}.
Matrices $\bgG$ and $\mathbf{E}$, together with constitutive relation (\ref{eq:StressIsotr}) and  fluxes definitions (\ref{eq:ThermaFluxIsotr}) and (\ref{eq:MassFluxIsotr}) are used to 
derive an explicit expression for the generalized amplitude vector $\mathbf{z}=(\mathbf{w}\, \mathbf{t})^{T}$, whose components are given by
\beq
\mathbf{z}(x_{2})=\left(\tilde{u}_1(x_2)\,\,  \tilde{u}_2(x_2)\,\,  \tilde{\theta}(x_2)\, \, \tilde{\eta}(x_2)\,\,  \tilde{\sigma}_{21}(x_2)\,\,  \tilde{\sigma}_{22}(x_2)\,\,  \tilde{q}_{2}(x_2)\,\,  \tilde{j}_{2}(x_2)\right)^{T}, 
\eeq
and then for the generalized solution $\mathbf{y}=\left(\mathbf{v}\,\, \mathbf{s}\right)^{T}=\mathbf{z} \,\textrm{exp}\left[i\left(\mathbf{k}\cdot\mathbf{x}-\omega t\right)\right]$. 
Vectors $\mathbf{z}$ and $\mathbf{y}$ assume, respectively, the form 
\beq
\mathbf{z}=\bgO\,\mathbf{E}\,\mathbf{a}, \quad \mathbf{y}=\bgO\,\mathbf{E}\,\mathbf{a}\, \textrm{exp}\left[i\left(\mathbf{k}\cdot\mathbf{x}-\omega t\right)\right], 
\label{ditrac}
\eeq
where the explicit expressions for the lines of the $8\times8$ matrix $\bgO$ are  
$$
\Omega_{1j}=\gamma_{j}^{(5)}, \ \ \Omega_{2j}=\gamma_{j}^{(6)}, \ \ \Omega_{3j}=\gamma_{j}^{(7)}, \ \   \Omega_{4j}=\gamma_{j}^{(8)},
$$
$$
\Omega_{5j}=G(\gamma_{j}^{(1)}+ik_1\gamma_{j}^{(6)}+ik_2\gamma_{j}^{(5)}), 
$$
$$
\Omega_{6j}=\frac{2G(1-\nu)}{1-2\nu}\gamma_{j}^{(2)}+\frac{2ik_1G\nu}{1-2\nu}\gamma_{j}^{(5)}+\frac{2ik_2G(1-\nu)}{1-2\nu}\gamma_{j}^{(6)}-\alpha\gamma_{j}^{(7)}-\beta\gamma_{j}^{(8)},
$$
\beq
\Omega_{7j}=-K(\gamma_{j}^{(3)}+ik_2\gamma_{j}^{(7)}), \ \ \Omega_{8j}=-D(\gamma_{j}^{(4)}+ik_2\gamma_{j}^{(8)}), \ \ \textrm{with}  \ \ j=1,\dots ,8.
\eeq
The second of \eq{ditrac} represents the formal generalized solution of the problem valid for  each $m^{th}$ layer  composing the  periodic cell of the laminate. Applying the transfer matrix method, 
equations \eq{ditrac} could be exploited  for studying the propagation and the
attenuation of oscillations induced by periodic boundary conditions on the whole multi-layered material.

\section*{Appendix B.  Recursive algorithm to determine the invariants of a characteristic polynomial  }
Eigenvalues of problem (\ref{eq:FBsyst}) are the roots of a characteristic polynomial $\mathcal{P}(\lambda)$ of the $8^{th}$ degree, which can be written in the form
\begin{equation}
\mathcal{P}(\lambda)=C_0+C_1\lambda+C_2 \lambda^2+ C_3 \lambda^3+C_4 \lambda^4+C_5 \lambda^5+C_6 \lambda^6+ C_7\lambda^7+ C_8 \lambda^8 
\label{eq:CharPol}
\end{equation}
The present Section describes a recursive method, called the Faddeev-LeVerrier algorithm \citep{Horst1935}, in order   to compute the invariants of characteristic polynomial (\ref{eq:CharPol}).
Coefficients $C_k$ of (\ref{eq:CharPol}) are recursively computed by means of the following formulas
\begin{subequations}
\begin{align}
&\mathbf{M}_0=\mathbf{0},
\hspace{0.3cm}
C_8=1
\hspace{0.2cm}
\textrm{at}
\hspace{0.1cm}
\textrm{step}
\hspace{0.1cm}
k=0,
\\
&\mathbf{M}_k=\mathbf{A}\mathbf{M}_{k-1}+C_{n-k+1}\mathbf{I},
\hspace{0.3cm}
C_{n-k}=-\frac{1}{k}\trace{\mathbf{A}\mathbf{M}_k}
\hspace{0.2cm}
\textrm{at}
\hspace{0.1cm}
\textrm{step}
\hspace{0.1cm}
k=1,...,8
\end{align}
\label{eq:FL-algorithm}
\end{subequations}
with matrix $\mathbf{A}=\mathbf{T}_{(1,n)}$ and $\mathbf{M}_k$ auxiliary matrices.
Applying equations (\ref{eq:FL-algorithm}) one finally has 
\begin{subequations}
\begin{align}
C_7&=-\trace{\mathbf{A}},
\label{eq:c7}
\\
C_6&=-\frac{1}{2}\trace{\mathbf{A}^2}+\frac{1}{2}\left(\trace{\mathbf{A}}\right)^2,
\label{eq:c6}
\\
C_5&=-\frac{1}{3}\trace{\mathbf{A}^3}
+\frac{1}{2}\trace{\mathbf{A}^2}\trace{\mathbf{A}}
-\frac{1}{6}\left(\trace{\mathbf{A}}\right)^3,
\label{eq:c5}
\\
C_4&=-\frac{1}{4}\trace{\mathbf{A}^4}+\frac{1}{3}\trace{\mathbf{A}}\trace{\mathbf{A}^3}+
\frac{1}{8}\left(\trace{\mathbf{A}^2}\right)^2-\frac{1}{4}\trace{\mathbf{A}^2}\left(\trace{\mathbf{A}}\right)^2+\frac{1}{24}\left(\trace{\mathbf{A}}\right)^4,
\label{eq:c4}
\\
C_3&=-\frac{1}{5}\trace{\mathbf{A}^5}+\frac{1}{4}\trace{\mathbf{A}}\trace{\mathbf{A}^4}+
\frac{1}{6}\trace{\mathbf{A}^2}\trace{\mathbf{A}^3}
-\frac{1}{6}\left(\trace{\mathbf{A}}\right)^2\trace{\mathbf{A}^3}\nonumber\\
&
-\frac{1}{8}\left(\trace{\mathbf{A}^2}\right)^2\trace{\mathbf{A}}
+\frac{1}{12}\left(\trace{\mathbf{A}}\right)^3\trace{\mathbf{A}^2}
-\frac{1}{120}\left(\trace{\mathbf{A}}\right)^5,
\label{eq:c3}
\\
C_2&=-\frac{1}{6}\trace{\mathbf{A}^6}
+\frac{1}{5}\trace{\mathbf{A}}\trace{\mathbf{A}^5}
+\frac{1}{8}\trace{\mathbf{A}^2}\trace{\mathbf{A}^4}
-\frac{1}{8}\trace{\mathbf{A}^4}\left(\trace{\mathbf{A}}\right)^2 
\nonumber\\
&
+\frac{1}{18}\left(\trace{\mathbf{A}^3}\right)^2
-\frac{1}{6}\trace{\mathbf{A}}\trace{\mathbf{A}^2}\trace{\mathbf{A}^3}
+\frac{1}{18}\trace{\mathbf{A}^3}\left(\trace{\mathbf{A}}\right)^3
-\frac{1}{48}\left(\trace{\mathbf{A}^2}\right)^3
\nonumber\\
&
+\frac{1}{16}\left(\trace{\mathbf{A}^2}\right)^2\left(\trace{\mathbf{A}}\right)^2 
-\frac{1}{48}\left(\trace{\mathbf{A}}\right)^4\trace{\mathbf{A}^2}
+\frac{1}{720}\left(\trace{\mathbf{A}}\right)^6,
\label{eq:c2}
\\
C_1&=-\frac{1}{7}\trace{\mathbf{A}^7}
+\frac{1}{6}\trace{\mathbf{A}}\trace{\mathbf{A}^6}
+\frac{1}{10}\trace{\mathbf{A}^2}\trace{\mathbf{A}^5}
-\frac{1}{10}\left(\trace{\mathbf{A}}\right)^2\trace{\mathbf{A}^5}
\nonumber\\
&
+\frac{1}{12}\trace{\mathbf{A}^3}\trace{\mathbf{A}^4}
-\frac{1}{8}\trace{\mathbf{A}}\trace{\mathbf{A}^2}\trace{\mathbf{A}^4}
+\frac{1}{24}\left(\trace{\mathbf{A}}\right)^3\trace{\mathbf{A}^4}
\nonumber\\
&
-\frac{1}{18}\trace{\mathbf{A}}\left(\trace{\mathbf{A}^3}\right)^2
-\frac{1}{24}\left(\trace{\mathbf{A}^2}\right)^2\trace{\mathbf{A}^3}
+\frac{1}{12}\left(\trace{\mathbf{A}}\right)^2\trace{\mathbf{A}^2}\trace{\mathbf{A}^3}
\nonumber\\
&
-\frac{1}{72}\left(\trace{\mathbf{A}}\right)^4\trace{\mathbf{A}^3}
+\frac{1}{48}\trace{\mathbf{A}}\left(\trace{\mathbf{A}^2}\right)^3
-\frac{1}{48}\left(\trace{\mathbf{A}}\right)^3\left(\trace{\mathbf{A}^2}\right)^2 
\nonumber\\
&
+\frac{1}{240}\left(\trace{\mathbf{A}}\right)^5\trace{\mathbf{A}^2}
-\frac{1}{5040}\left(\trace{\mathbf{A}}\right)^7,
\label{eq:c1}
\\
C_0&=-\frac{1}{8}\trace{\mathbf{A}^8}
+\frac{1}{7}\trace{\mathbf{A}}\trace{\mathbf{A}^7}
+\frac{1}{12}\trace{\mathbf{A}^2}\trace{\mathbf{A}^6}
-\frac{1}{12}\left(\trace{\mathbf{A}}\right)^2\trace{\mathbf{A}^6}
\nonumber\\
&
+\frac{1}{15}\trace{\mathbf{A}^3}\trace{\mathbf{A}^5}
-\frac{1}{10}\trace{\mathbf{A}}\trace{\mathbf{A}^2}\trace{\mathbf{A}^5}
+\frac{1}{30}\left(\trace{\mathbf{A}}\right)^3\trace{\mathbf{A}^5}
\nonumber\\
&
-\frac{1}{12}\trace{\mathbf{A}}\trace{\mathbf{A}^3}\trace{\mathbf{A}^4}
-\frac{1}{32}\left(\trace{\mathbf{A}^2}\right)^2\trace{\mathbf{A}^4}
+\frac{1}{16}\left(\trace{\mathbf{A}}\right)^2\trace{\mathbf{A}^2}\trace{\mathbf{A}^4}
\nonumber\\
&
-\frac{1}{96}\left(\trace{\mathbf{A}}\right)^4\trace{\mathbf{A}^4}
-\frac{1}{36}\trace{\mathbf{A}^2}\left(\trace{\mathbf{A}^3}\right)^2
+\frac{1}{36}\left(\trace{\mathbf{A}}\right)^2\left(\trace{\mathbf{A}^3}\right)^2
\nonumber\\
&
+\frac{1}{24}\trace{\mathbf{A}}\left(\trace{\mathbf{A}^2}\right)^2\trace{\mathbf{A}^3}
-\frac{1}{36}\left(\trace{\mathbf{A}}\right)^3\trace{\mathbf{A}^2}\trace{\mathbf{A}^3}
+\frac{1}{360}\left(\trace{\mathbf{A}}\right)^5\trace{\mathbf{A}^3}
\nonumber\\
&
+\frac{1}{384}\left(\trace{\mathbf{A}^2}\right)^4
-\frac{1}{96}\left(\trace{\mathbf{A}}\right)^2\left(\trace{\mathbf{A}^2}\right)^3
+\frac{1}{192}\left(\trace{\mathbf{A}^2}\right)^2\left(\trace{\mathbf{A}}\right)^4
\nonumber\\
&
-\frac{1}{1440}\left(\trace{\mathbf{A}}\right)^6\trace{\mathbf{A}^2}
+\frac{1}{32}\left(\trace{\mathbf{A}^4}\right)^2
+\frac{1}{40320}\left(\trace{\mathbf{A}}\right)^8
\label{eq:c0}
\end{align}
\label{eq:cCoefficients}
\end{subequations}
Since for a   $n^{th}$-degree characteristic polynomial, coefficient 
$C_0=(-1)^n \textrm{Det}(\mathbf{A})$, the 
Faddeev-LeVerrier algorithm
can also be exploited as a procedure to compute the determinant of a square matrix $\mathbf{A}$, which is usually a computationally expensive process. 
When matrix $\mathbf{A}$ is symplectic, as in the standard eigenvalue problem (\ref{eq:FBsyst}), the characteristic polynomial is palindromic \citep{Bronski2005}, meaning that  $\mathcal{P}(\lambda)=\sum_{j=0}^{2N}C_j\lambda^j$ with $C_{2N-j}=C_j$ and $N=4$.
It can be proved from equations (\ref{eq:cCoefficients})  that
 $C_8=C_0=1$, $C_7=C_1$, $C_6=C_2$ e $C_5=C_3$ and the $8^{th}$-degree polynomial $\mathcal{P}(\lambda)$, written as
\begin{equation}
\mathcal{P}(\lambda)=1+C_1\lambda+C_2 \lambda^2+ C_3 \lambda^3+C_4 \lambda^4+C_3 \lambda^5+C_2 \lambda^6+ C_1\lambda^7+  \lambda^8,
\label{eq:8thdegree_char_pol}
\end{equation}
results to be equivalent to the $4^{th}$-degree polynomial $\tilde{\mathcal{P}}(z)$
\begin{equation}
\tilde{\mathcal{P}}(z)=z^4+C_1 z^3+\left(C_2-4\right)z^2+\left(C_3-3C_1\right)z+
\left(C_4-2C_2+2\right),
\label{eq:4thdegreepolynomial}
\end{equation}
under conformal map $z=\lambda+\frac{1}{\lambda}$. 
Therefore, if $\lambda_k$ is the $k^{th}$ root for polynomial (\ref{eq:8thdegree_char_pol}), also $1/\lambda_k$ is a root for it.
Roots of polynomial (\ref{eq:4thdegreepolynomial}) can be analytically expressed.

\section*{Appendix C. Transfer matrix as power series of wave number $k_1$    }
When spatial damping (complex-valued wave vector $\mathbf{k}$ and real-valued angular frequency $\omega$) has to be investigated, transfer matrix $\mathbf{T}_m$ relative to the $m^{th}$ layer of the composite material introduced in equation (\ref{eq:y+New}), could be expressed as a power series of the wave number $k_1$.
Denoting  with $\mathbf{F}=\mathbf{M}^{-1}\mathbf{N}\ell_m$, matrix exponential $\textrm{exp}\left[\mathbf{F}\right]$, defined as
\begin{equation}
\textrm{exp}\left[\mathbf{F}\right]=\sum_{n=0}^{+\infty} \frac{1}{n!}\mathbf{F}^n,
\label{eq:ExponentialMatrixDefinition}
\end{equation}
is a function of the wave numbers $k_1$ and $k_2$, and of the angular frequency $\omega$, namely
$\textrm{exp}\left[\mathbf{F}\right]=f\left(k_1,k_2,\omega\right)$.
Based on expressions (\ref{mC}) and (\ref{eq:M&N}), matrix $\mathbf{F}$ can be decomposed as
\begin{equation}
\mathbf{F}=\mathbf{H}_0+k_1 \mathbf{H}_1+ k_1^2\mathbf{H}_2,
\end{equation}
where $\mathbf{H}_0$ collects terms that do not depend upon $k_1$,  $\mathbf{H}_1$ collects terms that linearly depend upon $k_1$, and  $\mathbf{H}_2$ collects terms that  depend upon $k_1^2$.
Matrix exponential $\textrm{exp}\left[\mathbf{F}\right]$ can therefore be expressed as
\begin{equation}
\textrm{exp}\left[\mathbf{F}\right]=\sum_{n=0}^{+\infty} \frac{1}{n!}\left(\mathbf{H}_0+k_1 \mathbf{H}_1+k_1^2\mathbf{H}_2\right)^n.
\end{equation}
Based upon the expression of the $n^{th}$ power of trinomial $\left(\mathbf{H}_0+k_1 \mathbf{H}_1+k_1^2\mathbf{H}_2\right)$, namely
\begin{eqnarray}
\left({\mathbf{H}}_0+k_1 \mathbf{H}_1+k_1^2 \mathbf{H}_2\right)^n&=&\sum_{r_1+r_2+r_3=n}n!\prod_{i=1}^3
\frac{\left(\mathbf{F}_{i-1}k_1^{i-1}\right)^{r_i}}{r_i!}=
\nonumber\\
&=&\sum_{j=0}^n\sum_{s=0}^{n-j}\frac{n!}{j!s!(n-j-s)!}\mathbf{H}_0^{n-j-s}\left(k_1\mathbf{H}_1\right)^s\left(k_1^2\mathbf{H}_2\right)^j,
\end{eqnarray}
equation (\ref{eq:y+New}) assumes the form
\begin{eqnarray}
\mathbf{y}_m^+&=&
\left(
\begin{array}{c c}
\mathbf{0} & \mathbf{I}\\
\mathbf{R} & i\mathbf{R}k_2+\mathbf{S}
\end{array}
\right)
\left[\sum_{n=0}^{+\infty}
\sum_{j=0}^n\sum_{s=0}^{n-j}\frac{1}{j!s!(n-j-s)!}\mathbf{H}_0^{n-j-s}\left(k_1\mathbf{H}_1\right)^s\left(k_1^2\mathbf{H}_2\right)^j
\right]
\nonumber\\
&
&
\left(
\begin{array}{c c}
\mathbf{0} & \mathbf{I}\\
\mathbf{R} & i\mathbf{R}k_2+\mathbf{S}
\end{array}
\right)^{-1}
\,\textrm{exp}\left[ik_2\ell_m\right]
\mathbf{y}_m^-.
\label{eq:y+NewApp2}
\end{eqnarray}
Consequently, transfer matrix $\mathbf{T}_m$ referred to the $m^{th}$ layer of the laminate, shows a polynomial dependence upon  wave number $k_1$ in the form
\begin{eqnarray}
\mathbf{T}_m&=&
\sum_{n=0}^{+\infty}
\sum_{j=0}^n\sum_{s=0}^{n-j}\frac{k_1^{s+2j}}{j!s!(n-j-s)!}
\left(
\begin{array}{c c}
\mathbf{0} & \mathbf{I}\\
\mathbf{R} & i\mathbf{R}k_2+\mathbf{S}
\end{array}
\right)
\mathbf{H}_0^{n-j-s}\,\mathbf{H}_1^s\, \mathbf{H}_2^j
\nonumber\\
&&
\left(
\begin{array}{c c}
\mathbf{0} & \mathbf{I}\\
\mathbf{R} & i\mathbf{R}k_2+\mathbf{S}
\end{array}
\right)^{-1}\,\textrm{exp}\left[ik_2\ell_m\right]
\end{eqnarray}
Transfer matrix of the entire unit cell $\mathbf{T}_{(1,n)}=\prod_{i=0}^{n-1}\mathbf{T}_{n-i}$, therefore, results to be expressed as a power series of $k_1$  and a suitable truncation of it can be employed in order to investigate wave propagation in the $\mathbf{e}_1$ direction.

\section*{Appendix D. Transfer matrix as power series of angular frequency $\omega$   }
In order to investigate temporal damping for the material of interest (complex-valued angular frequency $\omega$ and real-valued wave numbers $k_1$ and $k_2$), transfer matrix $\mathbf{T}_m$ introduced in equation (\ref{eq:y+New}) and relative to the $m^{th}$ material layer, could be expressed as a power series of the angular frequency $\omega$.
Referring to equation (\ref{eq:y+New}), and denoting with $\mathbf{F}=\mathbf{M}^{-1}\mathbf{N}\ell_m$, matrix exponential $\textrm{exp}\left[\mathbf{F}\right]$, defined as
\begin{equation}
\textrm{exp}\left[\mathbf{F}\right]=\sum_{n=0}^{+\infty} \frac{1}{n!}\mathbf{F}^n,
\label{eq:ExponentialMatrixDefinition}
\end{equation}
is a function of  wave numbers $k_1$ and $k_2$ and angular frequency $\omega$, namely $\textrm{exp}\left[\mathbf{F}\right]=f\left(k_1,k_2,\omega\right)$.
Based on expressions (\ref{mC}) and (\ref{eq:M&N}), matrix $\mathbf{F}$ can be decomposed as
\begin{equation}
\mathbf{F}=\mathbf{G}_0+\omega \mathbf{G}_1+\omega^2\mathbf{G}_2,
\end{equation}
collecting in $\mathbf{G}_0$ terms that do not depend upon $\omega$, in $\mathbf{G}_1$ terms that linearly depend upon $\omega$, and in $\mathbf{G}_2$ terms that  depend upon $\omega^2$.
Doing this, matrix exponential $\textrm{exp}\left[\mathbf{F}\right]$ results to be expressed as 
\begin{equation}
\textrm{exp}\left[\mathbf{F}\right]=\sum_{n=0}^{+\infty} \frac{1}{n!}\left(\mathbf{G}_0+\omega \mathbf{G}_1+\omega^2\mathbf{G}_2\right)^n.
\end{equation}
Since the $n^{th}$ power of trinomial $\left(\mathbf{G}_0+\omega \mathbf{G}_1+\omega^2\mathbf{G}_2\right)$ can be written as
\begin{eqnarray}
\left({\mathbf{G}}_0+\omega \mathbf{G}_1+\omega^2 \mathbf{G}_2\right)^n&=&\sum_{r_1+r_2+r_3=n}n!\prod_{i=1}^3
\frac{\left(\mathbf{G}_{i-1}\omega^{i-1}\right)^{r_i}}{r_i!}=\nonumber\\
&=&\sum_{j=0}^n\sum_{s=0}^{n-j}\frac{n!}{j!s!(n-j-s)!}\mathbf{G}_0^{n-j-s}\left(\omega\mathbf{G}_1\right)^s\left(\omega^2\mathbf{G}_2\right)^j,
\end{eqnarray}
one obtains that equation (\ref{eq:y+New}) is expressed in the form
\begin{eqnarray}
\mathbf{y}_m^+&=&
\left(
\begin{array}{c c}
\mathbf{0} & \mathbf{I}\\
\mathbf{R} & i\mathbf{R}k_2+\mathbf{S}
\end{array}
\right)
\left[\sum_{n=0}^{+\infty}
\sum_{j=0}^n\sum_{s=0}^{n-j}\frac{1}{j!s!(n-j-s)!}\mathbf{G}_0^{n-j-s}\left(\omega\mathbf{G}_1\right)^s\left(\omega^2\mathbf{G}_2\right)^j
\right]
\nonumber\\
&
&
\left(
\begin{array}{c c}
\mathbf{0} & \mathbf{I}\\
\mathbf{R} & i\mathbf{R}k_2+\mathbf{S}
\end{array}
\right)^{-1}
\,\textrm{exp}\left[ik_2\ell_m\right]
\mathbf{y}_m^-.
\label{eq:y+NewApp}
\end{eqnarray}
Transfer matrix $\mathbf{T}_m$ relative to the $m^{th}$ layer of the laminate, therefore, results to show a polynomial dependence upon angular frequency $\omega$, namely
\begin{eqnarray}
\mathbf{T}_m&=&
\sum_{n=0}^{+\infty}
\sum_{j=0}^n\sum_{s=0}^{n-j}\frac{\omega^{s+2j}}{j!s!(n-j-s)!}
\left(
\begin{array}{c c}
\mathbf{0} & \mathbf{I}\\
\mathbf{R} & i\mathbf{R}k_2+\mathbf{S}
\end{array}
\right)
\mathbf{G}_0^{n-j-s}\,\mathbf{G}_1^s\, \mathbf{G}_2^j
\nonumber\\
&&
\left(
\begin{array}{c c}
\mathbf{0} & \mathbf{I}\\
\mathbf{R} & i\mathbf{R}k_2+\mathbf{S}
\end{array}
\right)^{-1}\,\textrm{exp}\left[ik_2\ell_m\right].
\label{eq:TransMatrixPowerSeriesOmega}
\end{eqnarray}
From equation (\ref{eq:TransMatrixPowerSeriesOmega}), transfer matrix of the entire unit cell $\mathbf{T}_{(1,n)}=\prod_{i=0}^{n-1}\mathbf{T}_{n-i}$,  results to be expressed as a power series of $\omega$  and its truncation to a proper order can be exploited in order to investigate temporal damping.
\section*{Acknowledgments}
The authors acknowledge the financial support from National Group of Mathematical Physics (GNFM-INdAM).
MP would like to acknowledge financial support from the Italian Ministry of Education, University and Research (MIUR) to the research project of relevant national interest (PRIN 2017) ``XFAST-SIMS: Extra-fast and accurate simulation of complex structural systems'' (CUP: D68D19001260001).   

\bibliographystyle{elsarticle-harv}
\bibliography{SOFC_BF}


\end{document}